\documentclass[]{article}
\pdfoutput=1
\usepackage{jheppub}

\usepackage{setspace}
\usepackage{physics}
\usepackage{dsfont}
\usepackage{tensor}
\usepackage[normalem]{ulem}
\usepackage{xcolor}
\usepackage{graphicx}
\usepackage[export]{adjustbox}
\usepackage{mathtools}
\usepackage{mathbbol}
\usepackage[utf8]{inputenc}
\usepackage{amsmath}
\usepackage{bbold}
\usepackage{breqn}

\usepackage[title]{appendix}

\usepackage[english]{babel}
\usepackage{amssymb}
\usepackage{ragged2e}
\usepackage{etoolbox}
\usepackage{lipsum}
\usepackage{multirow}
\usepackage[most]{tcolorbox}
\usepackage{latexsym}
\usepackage{enumitem}
\usepackage[%
colorlinks=true,%
urlcolor=blue%
]{hyperref}

\usepackage{stackengine,scalerel}
\newcommand\dAlaux{%
  \Shortstack{\rule{12pt}{.6pt}\\
    \rule{.6pt}{10pt}\kern10pt\rule{1.4pt}{10pt}\\
    \rule{12pt}{1.4pt}}%
}
\newcommand\dAl{%
  \setstackgap{S}{0pt}%
  \setstackEOL{\\}%
  \scalerel*{\kern1pt\dAlaux\kern1pt}{\Delta}%
}

\usepackage{titlesec}
\usepackage{url}
\usepackage{caption}
\usepackage{subcaption}
\tcbset{
  highlight math/.style={notitle,nophantom,colframe=red,colback=yellow!25!white}
}

\newcommand*{\colorboxed}{}
\def\colorboxed#1#{%
  \colorboxedAux{#1}%
}
\newcommand*{\colorboxedAux}[3]{%
  \begingroup
    \colorlet{cb@saved}{.}%
    \color#1{#2}%
    \boxed{%
      \color{cb@saved}%
      #3%
    }%
  \endgroup
}

\usepackage{comment}
\usepackage{booktabs} 

\newcommand{\be}{\begin{equation}}
	\newcommand{\ee}{\end{equation}}
\newcommand{\beq}{\begin{equation}}
	\newcommand{\eeq}{\end{equation}}
\newcommand{\bea}{\begin{eqnarray}}
	\newcommand{\eea}{\end{eqnarray}}

\newcommand{\FTP} {\frac{\partial}{\partial \log \eps}}
\newcommand{\STP} {\frac{\partial}{\partial \log \eps} + \frac{1}{2} \frac{\partial^2}{(\partial \log \eps)^2}}

\newcommand{\eps}{{\epsilon}}
\newcommand{\apr}{\alpha'}
\newcommand{\EV}[1]{\left\langle #1 \right\rangle}

\newcommand{\AMR}[1]{\textcolor{purple}{[AMR: #1]}}

\newcommand{\RK}[1]{\textcolor{blue}{[RK: #1]}}

\title{A Worldsheet Derivation of the Classical Off-shell Boundary Action for the Dilaton in Half-Space}
   \author[a]{Amr Ahmadain,}
   \author[b]{Rifath Khan}

\affiliation[a] {Department of Physics, Swansea University, Swansea, SA2 8PP, UK}
\affiliation[b] {Stanford Institute for Theoretical Physics, 382 Via Pueblo, Stanford, CA 94305}
\emailAdd{amrahmadain@gmail.com}
\emailAdd{rifathkhantheo@gmail.com}

\abstract{We use the method of images to present a worldsheet derivation of the sphere partition function for the dilaton in half-space to leading order in $\apr$. We use Tseytlin's sphere prescription to obtain the total (bulk and boundary) off-shell classical bosonic string action for the dilaton in half-space and show that it satisfies the requirement for a well-defined variational principle with Dirichlet boundary conditions. 
}

\begin{document}

\maketitle

\section{Introduction}
The sphere partition function is a mysterious yet important object in string theory. Generally speaking, a worldsheet path integral calculation of the sphere partition function should produce the classical (tree-level) \textit{bulk} gravitational effective action, $I_{\text{bulk}}$, to all orders in $\alpha'$, in addition to all boundary terms, $I_{\text{bdy}}$, which are required to have a sound variational principle \cite{KT:2001}
\bea\label{eq:class_str_action}
I_{\text{sphere}}=I_{\text{bulk}}+I_{\text{bdy}}.
\eea
$I_{\text{bulk}}$ has been successfully derived from the worldsheet to leading order in $\apr$ \cite{TseytlinZeroMode1989,Andreev:1990iv,Ahmadain:2022eso} using Tseytlin's off-shell sphere prescription \cite{TSEYTLINMobiusInfinitySubtraction1988} and it was found to vanish on solutions of the equations of motion, i.e. on-shell. To date, as far as the authors know, there is no consistent way to derive or extract the boundary action $I_{\text{bdy}}$ from the worldsheet and Tseytlin's off-shell prescriptions are not successful either in doing that. 

This paper takes a very first step in that direction. We use Tseytlin's off-shell sphere prescription \cite{TSEYTLINMobiusInfinitySubtraction1988,Ahmadain:2022tew,Ahmadain:2022eso} and the method of images to evaluate the \textit{off-shell}  classical bosonic string action for the dilaton in half-space. We then show that the total (bulk plus boundary) off-shell action we derive satisfies the requirement for a well-defined variational principle with Dirichlet boundary conditions. 

Concretely, we take our target space $\mathcal{M}$ to be $\mathbb{R}_{+}\cross \mathbb{R}^{D-1}$. The total off-shell classical string action that we derive from the worldsheet is the sum of the bulk and boundary action for the dilaton
\begin{align}\label{eq:I_SphereIntro}
I_{\text{sphere}}&=-\frac{1}{2}\tilde{Z}_{\text{nz}} \apr \int_{\mathcal{M}} d^D Y  e^{-2\Phi}    \partial^2 \Phi  + \frac{1}{2}\apr\tilde{Z}_{\text{nz}}  \int_{\partial\mathcal{M}} d^{D-1} Y \ e^{-2\Phi} \partial_n \Phi\,, 
\end{align}
where $\partial_n \Phi$ is the normal derivative of $\Phi$ to the boundary. $I_{\text{bulk}}$ does not have a well-posed variational principle since it is second order in derivatives. $I_{\text{bdy}}$ was added by hand \cite{KT:2001} such that $I_{\text{sphere}}$ has a well-defined variational principle in target spaces with boundaries. In this paper, we however provide an exact first-principles worldsheet derivation of $I_{\text{bdy}}$. To be precise, by off-shell, we mean the nonconstant dilaton $\Phi(Y)$ is \textit{not} constrained to satisfy any equation of motion, bulk or boundary, during any step of our derivation. It is only in the last step that one imposes the equations of motion to get the classical on-shell action.


\subsection{Related Work}
Below, we give a quick overview of Tseytlin's off-shell prescription and summarize efforts to derive both terms in $I_{\text{sphere}}$ from the worldsheet. 

It is a standard result in string theory that the classical bulk action $I_{\text{bulk}}$ (tree-level cosmological constant), the one-point function (tree-level tadpole) and two-point function\footnote{Except for a string that comes in and back out without no interactions \cite{Erbin:2021}. In this case, there is an additional delta function in the numerator of the connected $\text{S}-$matrix that causes the two-function to diverge.} all vanish on-shell (on a string background defined by a CFT) in \textit{compact} target spacetime \cite{LiuPolchinski1988}. From the worldsheet perspective, the reason for this is simple. It is because after fixing the Diff and Weyl gauge symmetries of the genus-0 (sphere) worldsheet theory, one still has to divide by the infinite volume of the noncompact SL(2,$\mathbb{C}$). As a result, the on-shell sphere partition function with zero number of insertions vanishes
\begin{equation}\label{eq:Z_CFT}
-I_{\text{bulk}} = Z_{\text{CFT}} =\frac{1}{\text{SL(2,$\mathbb{C}$)}}\, Z_{\text{ghost}} \, Z_{\text{matter}}[\hat{g}] = \frac{K_0}{\infty}\,, 
\end{equation}
where $K_0$ is the genus-0 partition function without the CKG factor and $\hat{g}$ is a gauge-fixed metric. 

Within his first-quantized off-shell nonlinear sigma model (NSLM) formalism \cite{FT1,FT2,FT3}, Tseytlin made two proposals \cite{TSEYTLINMobiusInfinitySubtraction1988,TseytlinTachyonEA2001,APS-ClosedStringTachyons-2001} to obtain the \textit{bulk} classical off-shell action from the worldsheet and address the noncompactess of the SL(2,$\mathbb{C}$) group in the denominator of the sphere partition function. The regularized volume of $\operatorname{SL(2,\mathbb{C}})$ has a logarithmically divergent piece \cite{LiuPolchinski1988,TSEYTLINMobiusInfinitySubtraction1988,Ahmadain:2022tew,Eberhardt:2023lwd}. At the heart of both proposals, Tseytlin integrates over all $n$ vertex operator insertions and produces a logarithmically divergent term to cancel the log divergence of $\operatorname{SL(2,\mathbb{C}})$ in \eqref{eq:Z_CFT}. In the integrated vertex operator formalism, this directly leads to different types of divergences, logarithmic and power-law, in the $n$-point correlation function $K_{0,n}$ as $n$, $n-1$ or fewer operators approach each other in the integration region respectively. Regularizing these divergences require introducing a UV length cutoff $\eps$ on the worldsheet. To obtain the off-shell sphere partition function (the classical action), Tseytlin takes the derivative of the log of the UV cutoff with respect to $K_0$. This is Tseytlin's first prescription\footnote{The prescription \eqref{eq:T1} does not give the correct string vacuum for tachyons; it does not remove the tachyon tadpole. To deal with this problem, Tseytlin proposed his second prescription \cite{TseytlinSigmaModelEATachyons2001,TseytlinTachyonEA2001}
\be\label{eq:T2}
\qquad
-I_0^{\textbf{T2}} = Z_0^{\textbf{T2}} = \left(\STP\right) K_0.\quad\quad({\bf T2})\nonumber
\ee} 
\be\label{eq:T1}
\qquad\qquad\qquad
-I_{\text{bulk}} = Z_{\text{QFT}} = \FTP K_0.
\ee

In the NLSM, it is possible to find a renormalization group scheme where, to an arbitrary order in $\apr$, the QFT sphere partition function is defined to be 
\begin{equation}
K_0=V_{\text{gen}}\coloneqq \int_{\mathcal{M}} d^D Y e^{-2 \Phi}
\end{equation}
where $V_{\text{gen}}$ is the generalized volume factor  \cite{KT:2001,TseytlinPerelmanEntropy2007,Ahmadain:2022tew}.

In his first-quantized off-shell string theory formalism, Tseytlin breaks the sacred local conformal invariance of \textit{only} the matter part worldsheet NLSM. As a result, all calculable quantities, e.g. correlation functions, in the now worldsheet QFT will depend on the choice of the Weyl frame $\omega(z)$ when the worldsheet metric is expressed as $g_{a b}=e^{2 \omega(z)} \gamma_{a b}$. In QFT, breaking a local gauge symmetry is catastrophic since it immediately renders the underlying theory inconsistent. But it was shown in \cite{Ahmadain:2022tew} that infinitesimal changes in $\omega(z)$ on the worldsheet are equivalent to field redefinitions in target space, which corresponds to RG flow of the QFT. The conceptual foundations of Tseytlin's off-shell formalism and sphere prescriptions proposed in \cite{TSEYTLINMobiusInfinitySubtraction1988,TseytlinTachyonEA2001,APS-ClosedStringTachyons-2001} have been studied, analyzed and generalized in \cite{Ahmadain:2022tew, Ahmadain:2024hdp}.

This is the story in compact spacetime. However, in noncompact target spacetime (those with boundaries), there are subtleties \cite{Kraus:noncompactCFT:2002}. To begin with, it is not yet known how to use Tseytlin's off-shell sphere prescription \eqref{eq:Z_CFT} to obtain $I_{\text{bdy}}$ directly from the worldsheet \cite{KT:2001}. As far as the authors know, there is no consistent way yet to derive or extract the boundary action $I_{\text{bdy}}$ from the worldsheet. 

As we have discussed above, from the worldsheet perspective, the classical closed string off-shell action vanishes on-shell due to the infinite volume of $\operatorname{SL(2,\mathbb{C})}$. However, from the spacetime point of view, there is another reason why the classical string action is zero on-shell modulo a boundary term. It is because the dilaton in the classical action transforms in spacetime in such a way that only changes the overall normalization of the action. This property is also true for the closed string field theory action \cite{Erler:2022agw}. Although deriving the on-shell boundary action from the classical closed string field theory action is still a challenge (see section 10 in \cite{Sen:2024nfd}), some progress has recently been made in \cite{Firat:2024kxq} where the boundary terms of the closed string field action were identified at the massless level.

In \textit{noncompact} target spaces, the action is only stationary with respect to normalizable deformations of the CFT defining the on-shell string background but not for nonnormalizable deformations that survive at infinity. As pointed out in \cite{Erler:2022}, a constant shift of the dilaton is \textit{not} a variation of this form and thus, can receive nonzero boundary contributions from \textit{nonnormalizable} modes in a noncompact CFT \cite{Kraus:noncompactCFT:2002}. In the $\operatorname{AdS}_3$, this also leads to a non-zero one-point function \cite{troost2011ads3}. 

A peculiarity of the classical string action is that the generalized volume factor $e^{-2 \Phi}$ is the only term that depends on the nonconstant dilaton $\Phi(Y)$ and does not include derivatives 
\begin{equation}
I_{\text{bulk}} = \frac{1}{16 \pi G_N}\left[\int_{\mathcal{M}} d^D Y \sqrt{G} e^{-2 \Phi} \mathcal{L}(\partial_\mu \Phi, \phi_i) \right], 
\end{equation}
where \( \mathcal{L} \) depends \textit{only} on \( \partial_\mu \Phi \) and some other fields \( \phi_i \) (but not on \( \Phi \) itself). This feature of the classical action leads to an interesting consequence: the dilaton equation of motion implies the Lagrangian is a total derivative, and hence the on-shell action is a boundary term
\begin{equation}\label{eq:dilaton_td_term}
-2 e^{-2 \Phi} \mathcal{L} = \partial_\mu \left( e^{-2 \Phi} \frac{\delta \mathcal{L}}{\delta \partial_\mu \Phi} \right).
\end{equation}

For an asymptotically flat boundary, where gravity is weakly coupled, the classical string action to first order in $\apr$, including the Gibbons-Hawking-York boundary term \cite{York:1972sj,Gibbons:1976ue} is given by
\begin{equation}\label{eq:action_initial}
I_{\text{sphere}} = \frac{1}{16 \pi G_N}\left[\int_{\mathcal{M}} d^D Y\sqrt{G} e^{-2 \Phi}\left(-R-4(\nabla \Phi)^2+\cdots\right)-2 \int_{\partial \mathcal{M}} d^{D-1} Y \sqrt{h} e^{-2 \Phi} K\right],
\end{equation}
in which case $I_{\text{bdy}}$ becomes
\begin{equation}\label{eq:classical_action}
-I_{\text{bdy}}  = \frac{1}{8 \pi G_N} \int_{\partial \mathcal{M}} d^{D-1} Y \,\partial_n\left(e^{-2 \Phi} \sqrt{h}\right).
\end{equation}


The sphere and disk partition functions have been recently calculated for different backgrounds using methods different from those we use in this paper. The string partition function on the disk (for the open string) has been calculated in \cite{Eberhardt:diskPF:2021} where it was shown to give a finite nonzero value, contrary to the naive expectation that it vanishes due to the division by the infinite volume of the $\operatorname{PSL}(2, \mathbb{R})$ group \cite{LiuPolchinski1988}. The sphere partition function has been calculated in minimal models \cite{Mahajan:2021nsd} and shown to have unambiguous finite values and in two-dimensional quantum gravity \cite{Muhlmann:2021clm}. More recently, the sphere partition function for global $\operatorname{AdS}_3$ with pure NS flux has been studied \cite{eberhardt2023holographic} and found to match the conformal anomaly of the boundary holographic CFT as it should only in the supersymmeytric string. Because of the tachyon in bosonic string theory, a mismatch was found.

\subsection{Strategy and summary of results}\label{sec:Strategy}
In this paper, we use Tseytlin's sphere prescription to derive from the worldsheet the \textit{off-shell} classical boundary action for the dilaton in half-space (with a flat target spacetime metric) and obtain a total sphere action with a well-posed variational principle for Dirichlet boundary conditions.

We take our target space $\mathcal{M}$ to be the half-space $\mathbb{R}_{+}\cross \mathbb{R}^{D-1}$ with coordinates $X^{\mu} = (X^i,X^D)$ where $X^i \in (1, \cdots D-1)$ are coordinates along a codimension-1 wall at $X_D=0$. Strings in half-space is a $\mathbb{Z}_2$ orbifold of flat spacetime since $\mathbb{R}_{+}=\mathbb{R}/\mathbb{Z}_2$ \cite{Mertens:Thesis:2015}. In this half-space, the spherical worldsheet is \textit{restricted} to the right side of the wall (boundary) with coordinates $X^D \geq 0$ that splits the entire spacetime into two halves.

Our goal is to compute sphere partition function in half-space $Z_{\text{HS}}$ which is defined by restricting the path integral of embeddings of the sphere to the right side of the wall
\begin{equation}
    Z_{\text{HS}}[\Phi] = \int_{X^D\geq 0} [DX] e^{-I_{\text{NLSM}}[X,\Phi]}
\end{equation}
However, $Z_{\text{HS}}$ is not completely defined until we specify what happens to the spherical worldsheets when they bounce off the wall. To fully define $Z_{\text{HS}}$, one must take into account the contribution to the path integral coming from the bounce itself, analogous to what we do in the particle case \cite{Bastianelli:2006hq,Bastianelli:2008vh,Bastianelli:2009mw}, where the sign of the bounce decides the boundary condition imposed at the wall. In the string case, this is also what we do to specify boundary conditions in the worldsheet path integral. 

We use the method of images to evaluate it in a \textit{reflected} (doubled) space with a reflected dilaton $\tilde{\Phi}$ such that $Z_{\text{HS}}$ can be expressed as $\frac{1}{2}\, Z_{\text{RS}}$ 
\begin{align}
Z_{\text{HS}}[\Phi] &=\frac{1}{2}\, Z_{\text{RS}}[\tilde{\Phi}]  =\frac{1}{2}\int [DX] e^{-I_{\text{NLSM}}[X,\tilde{\Phi}]} 
\end{align}
where the dilaton $\tilde{\Phi}$ in reflected space is an infinitesimal deformation about the constant mode $\Phi_0$
\begin{align}\label{eq:reflected_dilaton}
\tilde{\Phi}(X^\mu) =  \Phi_0 + \phi(X^i,X^D) \Theta(X^D) + \phi(X^i,-X^D) \Theta(-X^D).
\end{align}

As in \cite{TseytlinZeroMode1989,Ahmadain:2022eso}, we use heat kernel regularization to regulate divergences in $Z_{\text{HS}}$ and $[\mathrm{D}X]$ by a inserting a factor of $e^{\epsilon \Delta}$ into divergent expressions, with $\Delta=-\nabla^2$.

We decompose the string into a zero mode $Y^{\mu}$ and a nonzero mode $\eta^{\mu}(z)$. Because of the kink in $\Phi(Y^D)$ at the wall, we find that $\EV{\abs{\eta^{D}(z)}}$ does not vanish as it always does in the NLSM expansion of deformations around \textit{compact} CFT with no target space boundary. We show that the near-wall (in terms of $\sqrt{\apr})$ part of $Z_{\text{HS}}$ is given by integral of the one-point function $\left\langle\abs{Y^D+\sqrt{\apr}\eta^D(z)}\right\rangle$ over the zero mode of the string $Y$ in target spacetime
\begin{align}
 Z_{\text{near}} &:=\frac{1}{2} Z_{\text{nz}} \int_{\frac{\abs{Y^D}}{\sqrt{\apr}}\sim 1} d^D Y  \left(-\frac{1}{4 \pi } \int d^2 z \sqrt{g}   R^{(2)} \left(\phi\left(Y^i, 0\right) +O(\apr)\right)\right) + Z_{\text{wall}} ,
\end{align}
where $Z_{\text{nz}}$ is the nonzero mode piece and $Z_{\text{wall}}$ is (after taking some appropriate limits) is given by 
\begin{align}
Z_{\text{wall}} &= - Z_{\text{nz}} \int_{\frac{\abs{Y^D}}{\sqrt{\apr}}\sim 1} d^D Y  \partial_D \phi\left(Y^i, 0\right) \left\langle\abs{Y^D+\sqrt{\apr}\eta^D(z)}\right\rangle  \\
&=-\frac{1}{2}\times 2  \tilde{Z}_{\text{nz}}\, \apr\int d^{D-1} Y \ e^{-2\Phi} \partial_D \Phi\left(Y^i, 0\right)     \left( \frac{ Y_c^2}{4} + \frac{- \ln \eps}{2} \right).
\end{align}\label{dilatonpre}

We then use Tseytlin's prescription \eqref{eq:T1} to obtain the classical boundary action for the dilaton in half-space 
\begin{equation}\label{eq:I_wall}
I_{\text{wall}} = -\FTP Z_{\text{wall}} 
=-\frac{\apr}{2}\tilde{Z}_{\text{nz}}  \int d^{D-1} Y \ e^{-2\Phi} \partial_D \Phi\left(Y^i, 0\right) = +\frac{\apr}{2}\tilde{Z}_{\text{nz}}  \int d^{D-1} Y \ e^{-2\Phi} \partial_n \Phi\left(Y^i, 0\right) ,
\end{equation}
where $\partial_n$ is the derivative in the direction of the outward-pointing normal $n^\mu$ i.e. $\partial_n=-\partial_D$.

Next, we again use Tseytlin's prescription to obtain the \textit{bulk} dilaton action in half-space from $Z_{\text{far}}$, which is the sphere partition function in the target spacetime region far from the wall

\begin{align}
I_{\text{bulk}} := I_{\text{far}} &= -\FTP Z_{\text{far}} 
=  -\frac{1}{2}\tilde{Z}_{\text{nz}} \apr \int_{\frac{Y^D}{\sqrt{\apr}}\gg 1} d^D Y  e^{-2\Phi}  \partial^2\Phi 
\end{align}

In spaces with boundaries, such as half-space, $I_{\text{bulk}}$ does \textit{not} have a well-defined variational principle, since varying it yields
\begin{align}
\delta I_{\text {bulk }}&=  \alpha^{\prime} \tilde{Z}_{\text{nz}} \int_{\frac{Y^D}{\sqrt{\apr}}\gg 1} d^D Y\left[e^{-2 \Phi} \partial^2 \Phi+\partial_\mu\left(e^{-2 \Phi} \partial^\mu \Phi\right)\right] \delta \Phi \nonumber \\
&- \frac{1}{2}\apr \tilde{Z}_{\text{nz}} \int_{\partial \mathcal{M}} d^{D-1} Y  e^{-2\Phi} \, \delta(\partial_n \Phi)  - \alpha^{\prime} \tilde{Z}_{\text{nz}} \int_{\partial \mathcal{M}} d^{D-1} Y e^{-2 \Phi} \partial_n \Phi \, \delta \Phi . \label{eq:deltaI_bulk}
\end{align}
The first boundary term in $\delta I_{\text{bulk}}$ spoils the variational principle. It is the Gibbons-Hawking-York (GHY)-like term of the free two-derivative action for $\Phi$ which must be accounted for, i.e. by adding an \textit{additional} boundary term whose variation exactly cancels with it to have a well-posed variational principle. This additional boundary term is $I_{\text{wall}}$.

The total classical string action in half-space, that we derive from the worldsheet in this paper, is the sum of the bulk and boundary parts for the dilaton
\begin{align}
I_{\text{sphere}}:= I_{\text{HS}}&=I_{\text{bulk}}+I_{\text{wall}} \nonumber\\
&=-\frac{1}{2}\tilde{Z}_{\text{nz}} \apr \int_{\frac{Y^D}{\sqrt{\apr}}\gg 1} d^D Y  e^{-2\Phi}    \partial^2 \Phi  +\frac{1}{2} \apr\tilde{Z}_{\text{nz}}  \int d^{D-1} Y \ e^{-2\Phi} \partial_n \Phi\left(Y^i, 0\right) .
\label{eq:I_HS}
\end{align}

Varying $I_{\text{sphere}}$ now gives
\begin{align}
\delta I_{\text {sphere}} =&\, \delta I_{\text {bulk }} + \delta I_{\text {wall}} \nonumber  \\
&= \alpha^{\prime} \tilde{Z}_{\text{nz}} \int_{\frac{Y^D}{\sqrt{\apr}}\gg 1} d^D Y\left[e^{-2 \Phi} \partial^2 \Phi+\partial_\mu\left(e^{-2 \Phi} \partial^\mu \Phi\right)\right] \delta \Phi -2 \alpha^{\prime} \tilde{Z}_{\text{nz}} \int_{\partial \mathcal{M}} e^{-2 \Phi} \partial_n \Phi \, \delta \Phi.
\end{align}
Therefore, we observe that $I_{\text {sphere}}$ \eqref{eq:I_HS} has a well-defined variational principle consistent with Dirichlet boundary conditions.  



\textbf{Plan of paper:} In section \ref{sec:PF_halfspace}, we present the conceptual foundations of calculating $Z_{\text{HS}}$ and $I_{\text{HS}}$ using the method of images in analogy to how boundary conditions are imposed in the worldline case.

In section \ref{sec:Setup}, we setup the calculation that we do in section \ref{sec:BoundaryAction}. In appendix \ref{sec:coincorrfunc}, we present the details of computing the one-point function $\left\langle\abs{Y^D+\sqrt{\apr}\eta^D(z)}\right\rangle$. We first expand $\eta^D(z)$ and the Laplacian in real spherical harmonics and then constrain the path integral by inserting a delta function to compute $\left\langle\abs{Y^D+\sqrt{\apr}\eta^D(z)}\right\rangle$.

In section \ref{sec:BoundaryAction}, we finally put things together to arrive at the central result of this work. We  write down the expression for $I_{\text{sphere}}$, the total off-shell classical action in half-space and demonstrate that it has a well-posed variational principle for Dirichlet boundary conditions.

Finally, in section \ref{sec:discussion}, we make the observation that the expectation value of $\abs{\eta^D(z)^N}$ corresponds to $N$-th moment of the half-normal distribution for the random variable $Y=|X|$. We then argue, albeit speculatively, that $\left\langle\abs{\sqrt{\apr}\eta^D(z)}\right\rangle$  is related to the expectation value of the so-called boundary local time $\EV{L^0_t}$ in reflected Brownian motion. We conclude with an outlook on future directions and open questions.

\section{The sphere partition function in $\mathbb{R}_{+}\cross \mathbb{R}^{D-1}$}\label{sec:PF_halfspace}
Let our target space $\mathcal{M}$ be the half-space $\mathbb{R}_{+}\cross \mathbb{R}^{D-1}$ with coordinates $X^{\mu}$. A co-dimension one wall (boundary) sits at $X^D=0$. In this half-space, the spherical worldsheet will be \textit{restricted} to the right side of the wall with coordinates $X^D \geq 0$.

Consider the worldsheet NLSM path integral in $\mathbb{R}_{+}\cross \mathbb{R}^{D-1}$ given by the off-shell sphere partition function
\begin{equation}\label{eq:Z_HS_section2}
    Z_{\text{HS}}[G,\Phi,\omega] = \int_{X^D\geq0}  [DX] \ e^{-I_{\text{NLSM}}[X,G,\Phi,\omega,\gamma]},
\end{equation}
where $\gamma$ is the worldsheet conformal metric, $\omega$ is the worldsheet conformal factor, and $X^\mu$ are the embedding fields on the worldsheet. For details of gauge fixing (properly choosing $\gamma$ and $\omega$) in off-shell sphere partition functions, see \cite{Ahmadain:2022tew}. 

In this work, we take the target spacetime metric to be flat, and the dilaton field is expanded about a constant mode\footnote{The mass dimensions of the metric and dilaton are $[G]=0$, $[\Phi] = 0$, and $[\apr]=-2$.}
\begin{equation}
    G_{\mu\nu}(X)=\delta_{\mu\nu}, \quad  \Phi(X) = \Phi_0 + \phi(X), \quad   \text{and} \quad  \phi(X) \ll 1
\end{equation}
Here, the function $\Phi(X)$ is defined \textit{only} for $X^D\geq0$, i.e, on the right side of the wall.

If we could find a way evaluate $Z_{\text{HS}}$ with the restriction $X^D\geq0$ i.e. restricting the worldsheet to the right side of the wall, then we could obtain the classical boundary action for the dilaton. Evaluating the worldsheet NLSM path integral with this restriction to half-space is highly non-trivial. To do this, we combine the methods of off-shell string theory studied in \cite{Ahmadain:2022tew,Ahmadain:2022eso} with the method of images to evaluate the NLSM worldsheet string path integral in $\mathbb{R}_{+}\cross \mathbb{R}^{D-1}$. 



\subsection{The method of images}\label{sec:partion_fun_ref_st}

\begin{figure*}
    \centering
    \includegraphics[width=1\linewidth]{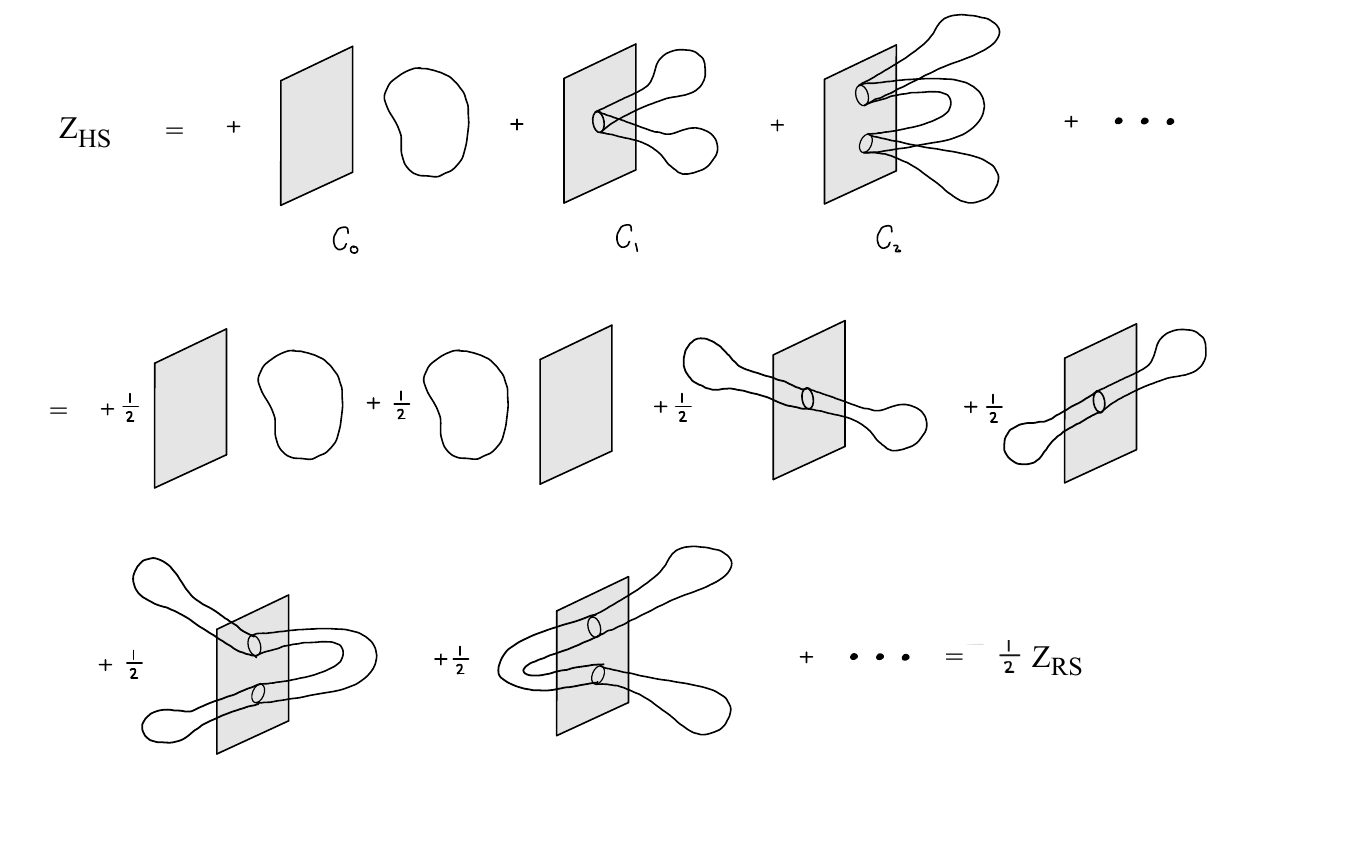}
\caption{Method of images in target spacetime with Neumann boundary conditions. The partition function in half-space $Z_{\text{HS}}$ is one-half its counterpart in reflected (doubled) space $Z_{\text{RS}}$.}
    \label{fig:method of images}
\end{figure*}

The quantity we want calculate is
\begin{align}
Z_{\text{HS}} &= \int_{X^D\geq 0} [DX] e^{-I_{\text{NLSM}}[X,\Phi,\omega,\gamma]}.
\end{align}
This is the path integral of embeddings of the spherical worldsheet in half-space i.e. on the right side of the wall ($X^D \geq 0$). Let $\mathcal{C}$ be the set of all embeddings of the worldsheet on the right side of the wall. We aim to classify this set by the number of bounces of the worldsheet off the wall. An example of an element of this set is an embedding for which no points on the worldsheet touch the wall (i.e $X^D>0$ everywhere on the worldsheet). Denote this class of embeddings by $\mathcal{C}_0$ (see fig. \ref{fig:method of images}). There is also a class of embeddings in which the worldsheet bounces off the wall \textit{once}, by which we mean that the worldsheet intersects the wall in one circle, without passing through it. Denote this class by $\mathcal{C}_1$. The worldsheet can also bounce twice off the wall, i.e. the worldsheet intersects the wall in two circles. Denote the class of all embeddings based on the number of bounces $n$ off the wall by $\mathcal{C}_n$. 

However, this is not all of the embeddings one can consider. For example, one can also consider an embedding which intersects the wall in only one point or in a disc. We expect all such embeddings to be a measure zero set in the path integral and thus can be thrown away. Therefore, all embeddings that give a nonzero contribution can be classified by the number of bounces off the wall.

It is important to note that whenever the worldsheet bounces off the wall, there is a contribution to the path integral. This is analogous to the particle case, where the contribution of the bounce is a plus sign for Neumann boundary conditions and a minus sign for Dirichlet boundary conditions \cite{Bastianelli:2006hq,Bastianelli:2008vh,Bastianelli:2009mw}. We expect that the sign in the string bounce case should impose some boundary condition, but it is unclear which boundary condition it is meant to impose (since it was only an analogy). We will choose the plus sign as the contribution from the worldsheet bounce, and find that it gives results consistent with Dirichlet boundary conditions. A deeper understanding of why this sign imposes Dirichlet boundary conditions is needed, and we leave this as an open problem.

Consider the embedding that bounces off the wall once. The worldsheet intersects the wall in a circle. The action being an integral on the worldsheet can be split into two pieces, each for a piece emanating from the circle. Now, if we reflect the spacetime and a single piece, the action remains the same. So we get an embedding in the reflected spacetime that crosses the wall once. But, there are exactly two embeddings in the reflected target space that correspond to a single embedding in half space. The same argument holds for any number of bounces (including zero bounces). Therefore, we can express the path integral of embeddings in half-space as one half the path integral in the reflected space. We will explore in a bit more detail the relationship between half-space and reflected space in light of reflected Brownian motion and stochastic dynamics in section \ref{sec:discussion}.

\begin{align}
Z_{\text{HS}} &= \int_{X^D\geq 0} [DX] e^{-I_{\text{NLSM}}[X,\Phi,\omega,\gamma]}\\ \nonumber
&=\frac{1}{2}\, Z_{\text{RS}} = \int \frac{1}{2} [DX] e^{-I_{\text{NLSM}}[X,\tilde{\Phi},\omega,\gamma]}
\end{align}

Next, we explain the set-up of the calculation of $Z_{\text{RS}}$.

\section{Setup of calculation}\label{sec:Setup}
In terms of the NLSM worldsheet action, both the metric and dilaton are reflected \textit{about} the wall to the other side. The spacetime metric is flat \textit{everywhere}, but the dilaton in this reflected space $\tilde{\Phi}(X^\mu)$ is defined for $X^D \in \mathbb{R}$. More concretely, $\tilde{\Phi}(X^\mu)$ is the sum of two terms, each takes values on one side of the wall
\begin{align} \label{eq:extended_dilaton}
    \tilde{\Phi}(X^\mu) = \Phi_0+\tilde{\phi}(X^\mu) &=  \Phi(X^i,X^D) \Theta(X^D) + \Phi(X^i,-X^D) \Theta(-X^D)\\ 
    &= \Phi_0 + \phi(X^i,X^D) \Theta(X^D) + \phi(X^i,-X^D) \Theta(-X^D) \nonumber.
\end{align}

To carry out the NLSM expansion, we split the string $X^\mu$ into a zero mode $Y^\mu$ and a non-zero mode $\eta^\mu(z)$
\begin{equation}
    X^\mu(z) = Y^\mu + \sqrt{\apr} \eta^\mu(z).
\end{equation}

The deformed NLSM action is given by
\begin{align}
    I_{\text{NLSM}} &= \frac{1}{4 \pi \apr} \int d^2 z \sqrt{g}\left(g^{ab}\partial_a X^\mu \partial_b X^\nu  \delta_{\mu \nu} +  \apr R^{(2)} \tilde{\Phi}(X)  \right)\\
    &= \frac{1}{4 \pi \apr} \int d^2 z \sqrt{g}\left(\apr g^{ab}\partial_a \eta^\mu \partial_b \eta^\nu  \delta_{\mu \nu} +  \apr R^{(2)} \tilde{\Phi}\left(Y+\sqrt{\apr}\eta\right)  \right)\\
    &= \frac{1}{4 \pi \apr} \int d^2 z \sqrt{g}\left(\apr g^{ab}\partial_a \eta^\mu \partial_b \eta^\nu  \delta_{\mu \nu} +  \apr R^{(2)} \Phi_0  \right) +\frac{1}{4 \pi \apr} \int d^2 z \sqrt{g}\left(  \apr R^{(2)} \tilde{\phi}(Y+ \sqrt{\apr} \eta)  \right)\\
    &= \frac{1}{4 \pi \apr} \int d^2 z \sqrt{g}\left(\apr g^{ab}\partial_a \eta^\mu \partial_b \eta^\nu  \delta_{\mu \nu}  \right) + 2 \Phi_0 +\frac{1}{4 \pi \apr} \int d^2 z \sqrt{g} \left(  \apr R^{(2)} \tilde{\phi}(Y+\sqrt{\apr} \eta)  \right)
    \\
    &=I_{\text{P}}[\eta,g] + V[\eta,g,\tilde{\phi}],
\end{align}
where 
\begin{equation}
V[\eta,g,\tilde{\phi}] = \frac{1}{4 \pi } \int d^2 z \sqrt{g}   R^{(2)} \tilde{\phi}(Y+ \sqrt{\apr}\eta(z)).
\end{equation}

The path integral measure is given by
\begin{align}
    [DX] = d^D Y [ D\sqrt{\apr}\eta] \,.
\end{align}
and the matter partition function in half-space becomes
\begin{align}\label{eq:Z_HS_section4}
 Z_{\text{HS}}&=\frac{1}{2} \int [DX] e^{-I_{\text{NLSM}}[X,\tilde{\Phi},\omega,\gamma]},\\&= \frac{1}{2}\int d^D Y \int [D\sqrt{\apr}\eta] e^{-I_{\text{p}}[\eta,g]} e^{- V[\eta,g,\tilde{\phi}]}\\&= \frac{1}{2}\int d^D Y \int [D\sqrt{\apr}\eta] e^{-I_{\text{p}}[\eta,g]} \left(1-V+O(V^2)\right)\\&=\frac{1}{2} Z_{\text{nz}} \int d^D Y  \left(1-\left\langle V\right\rangle+ \left\langle O(V^2)\right\rangle\right),
\end{align}
where $Z_{\text{nz}} := \int \left[D\sqrt{\apr}\eta\right] e^{-I_{\text{p}}[\eta,g]}$ is the path integral for the non-zero modes and $\left\langle V\right\rangle$ is defined by 
\begin{equation}
\left\langle V\right\rangle =\frac{1}{Z_{\text{nz}}} \int [D\sqrt{\apr}\eta] e^{-I_{\text{p}}[\eta,g]} V \,.
\end{equation}
Since $\phi \ll 1$, the higher order terms in $V$ are suppressed, so we can drop the $O(V^2)$ terms.

Now we compute $\left\langle V\right\rangle$.
\begin{align}
    \left\langle V\right\rangle &= \frac{1}{4 \pi } \int d^2 z \sqrt{g}   R^{(2)} \int [D\sqrt{\apr}\eta]\, e^{-I_{\text{p}}[\eta,g]}  \tilde{\phi}(Y^{\mu}+\sqrt{\apr}\eta^{\mu}(z)) \\
    &= \frac{1}{4 \pi } \int d^2 z \sqrt{g}   R^{(2)} \left\langle \tilde{\phi}(Y^{\mu}+\sqrt{\apr}\eta^{\mu}(z))\right\rangle.
\end{align}
If $Y^{\mu}$ is far away from the wall in string units, i.e. $\frac{\abs{Y^D}}{\sqrt{\apr}} \gg 1$, then one can Taylor expand $ \tilde{\phi}(X^{\mu})$  about the point $Y^{\mu}$

\begin{align}
    \left\langle \tilde{\phi}\left(Y^{\mu}+\sqrt{\apr}\eta^{\mu}(z)\right)\right\rangle &= \tilde{\phi}\left(Y^{\mu}\right) \left\langle \mathbf{1} \right\rangle + \sqrt{\apr}\partial_\mu \tilde{\phi}\left(Y^{\mu}\right) \left\langle \eta^\mu (z) \right\rangle  + \frac{\apr}{2} \partial_\mu \partial_\nu \tilde{\phi}\left(Y^{\mu}\right) \left\langle \eta^\mu(z) \eta^\nu (z) \right\rangle + O(\alpha^{\prime 3/2}).
\end{align}
Here, $\left\langle \mathbf{1} \right\rangle=1$ and the one-point function vanishes (because it an odd integral). Then
\begin{align}
    \left\langle \tilde{\phi}\left(Y^{\mu}+\sqrt{\apr}\eta^{\mu}(z)\right)\right\rangle &= \tilde{\phi}\left(Y^{\mu}\right) + \frac{\apr}{2} \partial_\mu \partial_\nu \tilde{\phi}\left(Y^{\mu}\right) \left\langle \eta^\mu(z) \eta^\nu (z) \right\rangle + O(\alpha^{\prime 3/2}).
\end{align}
It may be worth pointing out that this far-from-the-wall region where the worldsheet path integral is \textit{not} restricted is the case considered in \cite{Ahmadain:2022eso}, and this is the reason why there is no notion of a one-point to begin with.

However, if $Y^\mu$ is close to the wall, i.e $\frac{\abs{Y^D}}{\sqrt{\apr}} \sim 1$, then the \textit{kink} in $\tilde{\phi}\left(X^i, X^D\right)$ at the wall does matter. In this region near the wall, the dilaton will approximately be a linear function of the absolute value of $X_D$ on either side of the wall
\begin{align}
\tilde{\phi}\left(X^i, X^D\right) \approx \phi\left(X^i, 0\right) + \partial_D \phi\left(X^i, 0\right) \abs{X^D}\,,
\end{align}
where $\partial_D \phi\left(X^i, 0\right)$ is the right limit of the first derivative (as strictly speaking, the derivative doesn't exist on the wall due to the kink):
\begin{equation}
\partial_D \phi\left(X^i, 0\right) := \lim_{X^D \to 0^+} \partial_D \phi\left(X^i, X^D\right).
\end{equation}
Now we can Taylor expand along the tangential direction to the wall about the point $Y^i$
\begin{align}
\tilde{\phi}\left(Y^i+\sqrt{\apr}\eta^i(z), Y^D+\sqrt{\apr}\eta^D(z)\right) &\approx \phi\left(Y^i+\sqrt{\apr}\eta^i(z), 0\right) + \partial_D \phi\left(Y^i+\sqrt{\apr}\eta^i(z), 0\right) \abs{Y^D+\sqrt{\apr}\eta^D(z)}\\
&=\phi\left(Y^i, 0\right)+ \sqrt{\apr} \partial_i \phi\left(Y^i, 0\right) \eta^i(z) \\
&+\partial_D \phi\left(Y^i, 0\right) \abs{Y^D+\sqrt{\apr}\eta^D(z)}\\
&+\sqrt{\apr} \partial_i\partial_D \phi\left(Y^i, 0\right) \eta^i(z)\abs{Y^D+\sqrt{\apr}\eta^D(z)} +O(\apr).
\end{align}
The expectation value of all terms odd in $\eta^i(z)$ vanishes in a free theory. Hence, 
\begin{align}
    \left\langle \tilde{\phi}\left(Y^\mu+\sqrt{\apr}\eta^\mu(z)\right)\right\rangle &= \phi\left(Y^i, 0\right) +\partial_D \phi\left(Y^i, 0\right) \left\langle\abs{Y^D+\sqrt{\apr}\eta^D(z)}\right\rangle +O(\apr).
\end{align}
Thus, it is the near-wall region that leads to a non-zero one-point function in the limit ${X^D \to 0^+}$ because the wiggling part of the string $\eta^D(z)$ is restricted to half-space by the presence of the wall. 

An explicit calculation of the exact one-point function $\left\langle\abs{Y^D+\sqrt{\apr}\eta^D(z)}\right\rangle$ is presented in appendix \ref{sec:coincorrfunc}.



\section{The classical boundary dilaton action in half-space }\label{sec:BoundaryAction}
In this section, we put things together to compute the total classical off-shell action in half-space, $I_{\text{HS}}$, and show it obeys a well-defined variational principle for Dirichlet boundary conditions. We first use the result in appendix \ref{sec:coincorrfunc} for the one-point $\left\langle \abs{Y^D+\sqrt{\apr}\eta^D(z_0)}\right\rangle$ to write down an expression for the boundary dilaton action from the near-wall spacetime region. We then compute the bulk dilaton action $I_{\text{bulk}}$, i.e. the dilaton kinetic term, from the far-wall region. This bulk action comes with its own boundary contribution \cite{Ahmadain:2022eso}. To obtain the total action, we add all bulk and boundary contributions and apply Tsyetlin's prescriptions. We then find that the two boundary terms cancel out (as they should) and we are left with an off-shell action for the dilaton in half-space that has a well-defined variational principle for a Dirichlet boundary condition.

In appendix \ref{sec:coincorrfunc}, we show that 
\begin{align}\label{modoneptexpvalue}
\left\langle \abs{Y^D+\sqrt{\apr}\eta^D(z_0)}\right\rangle
 &=\sqrt{\apr}\left(\frac{\sqrt{\Omega}}{\sqrt{\pi}}   e^{-\frac{(Y^D)^2}{\apr \Omega}} + \frac{Y^D}{\sqrt{\apr}}  \text{erf}\left(\frac{Y^D}{\sqrt{\apr \Omega}}\right)\right) \\
 &= \sqrt{\apr}\left(\left \langle\abs{\eta^D(z_0)}\right\rangle e^{-\frac{(Y^D)^2}{\apr \Omega}} + \frac{Y^D}{\sqrt{\apr}}  \text{erf}\left(\frac{Y^D}{\sqrt{\apr \Omega}}\right)\right) .
\end{align}

The sphere partition function in half-space \eqref{eq:Z_HS_section4} splits into two pieces, each in a different spacetime region: near the wall and far from the wall
\begin{align} \label{eq:Z_HS}
 Z_{\text{HS}}
=\frac{1}{2}Z_{\text{nz}} V_{Y^D} + Z_{\text{near}}+ Z_{\text{far}} + O(V^2)\,,
\end{align}
where $V_{Y^D}=\int d^D Y$ and
\begin{align}
Z_{\text{near}} &:=\frac{1}{2} Z_{\text{nz}} \int_{\frac{\abs{Y^D}}{\sqrt{\apr}}\sim 1} d^D Y  \left(-\frac{1}{4 \pi } \int d^2 z \sqrt{g}   R^{(2)} \left\langle \tilde{\phi}(Y^{\mu}+\sqrt{\apr}\eta^{\mu}(z))\right\rangle\right).\\  
Z_{\text{far}}&:=\frac{1}{2}Z_{\text{nz}} \int_{\frac{\abs{Y^D}}{\sqrt{\apr}}\gg 1} d^D Y  \left(-\frac{1}{4 \pi } \int d^2 z \sqrt{g}   R^{(2)} \left\langle \tilde{\phi}(Y^{\mu}+\sqrt{\apr}\eta^{\mu}(z))\right\rangle\right).
\end{align}
Far from the wall, $Z_{\text{far}}$ gives the bulk contribution to the action while, due to the presence of the wall, $Z_{\text{near}}$ gives an additional contribution which is given by 
\begin{align}\label{eq:dilaton_bdy_matter}
 Z_{\text{near}} &:= \frac{1}{2}Z_{\text{nz}} \int_{\frac{\abs{Y^D}}{\sqrt{\apr}}\sim 1} d^D Y  \left(-\frac{1}{4 \pi } \int d^2 z \sqrt{g}   R^{(2)} \left(\phi\left(Y^i, 0\right) +\partial_D \phi\left(Y^i, 0\right) \left\langle\abs{Y^D+\sqrt{\apr}\eta^D(z)}\right\rangle  +O(\apr)\right)\right)\\ 
 &=\frac{1}{2} Z_{\text{nz}} \int_{\frac{\abs{Y^D}}{\sqrt{\apr}}\sim 1} d^D Y  \left(-\frac{1}{4 \pi } \int d^2 z \sqrt{g}   R^{(2)} \left(\phi\left(Y^i, 0\right) +O(\apr)\right)\right) + Z_{\text{wall}}\,, \label{blehblah}
\end{align}
where $Z_{\text{wall}}$ is 
\begin{align}
  Z_{\text{wall}} &=\frac{1}{2}  Z_{\text{nz}} \int_{\frac{\abs{Y^D}}{\sqrt{\apr}}\sim 1} d^D Y  \left(-\frac{1}{4 \pi } \int d^2 z \sqrt{g}   R^{(2)}  \partial_D \phi\left(Y^i, 0\right) \left\langle\abs{Y^D+\sqrt{\apr}\eta^D(z)}\right\rangle  \right)\\
  &= \frac{1}{2} Z_{\text{nz}} \int_{\frac{\abs{Y^D}}{\sqrt{\apr}}\sim 1} d^D Y  (-2)  \partial_D \phi\left(Y^i, 0\right) \left\langle\abs{Y^D+\sqrt{\apr}\eta^D(z)}\right\rangle  \\
  &=-\frac{1}{2}\times 2  Z_{\text{nz}} \int d^{D-1} Y \  \partial_D \phi\left(Y^i, 0\right)    \int_{\frac{\abs{Y^D}}{\sqrt{\apr}}\sim 1} d Y^D    \sqrt{\apr}\left(\frac{\sqrt{\Omega}}{\sqrt{\pi}}   e^{-\frac{(Y^D)^2}{\apr \Omega}} + \frac{Y^D}{\sqrt{\apr}}  \text{erf}\left(\frac{Y^D}{\sqrt{\apr \Omega}}\right)\right) .\label{dilatonpre}
\end{align}
The integrand in \eqref{dilatonpre} is valid \textit{only} near the wall where $Y^D \sim \sqrt{\apr}$ and thus we integrate it from $-\frac{Y_c \sqrt{\apr}}{2}$ to $\frac{Y_c \sqrt{\apr}}{2}$ where the cutoff $Y_c$ is a dimensionless constant.

\begin{align}
  Z_{\text{wall}} &=-\frac{1}{2}\times 2  Z_{\text{nz}} \int d^{D-1} Y \  \partial_D \phi\left(Y^i, 0\right) \int_{-\frac{Y_c \sqrt{\apr}}{2}}^{\frac{Y_c \sqrt{\apr}}{2}} d Y^D    \sqrt{\apr}\left(\frac{\sqrt{\Omega}}{\sqrt{\pi}}   e^{-\frac{(Y^D)^2}{\apr \Omega}} + \frac{Y^D}{\sqrt{\apr}}  \text{erf}\left(\frac{Y^D}{\sqrt{\apr \Omega}}\right)\right)\\
  &=-\frac{1}{2}\times 2  Z_{\text{nz}} \int d^{D-1} Y \  \partial_D \phi\left(Y^i, 0\right) 2\int_{0}^{\frac{Y_c \sqrt{\apr}}{2}} d Y^D    \sqrt{\apr}\left(\frac{\sqrt{\Omega}}{\sqrt{\pi}}   e^{-\frac{(Y^D)^2}{\apr \Omega}} + \frac{Y^D}{\sqrt{\apr}}  \text{erf}\left(\frac{Y^D}{\sqrt{\apr \Omega}}\right)\right)\\
  &=-\frac{1}{2}\times 2  Z_{\text{nz}} \int d^{D-1} Y \  \partial_D \phi\left(Y^i, 0\right)  \frac{\apr}{4}  \left( \frac{2 e^{-\frac{Y_c^2}{4 \Omega}} Y_c \sqrt{\Omega}}{\sqrt{\pi}} + (Y_c^2 + 2 \Omega) \text{erf}\left(\frac{Y_c}{2 \sqrt{\Omega}}\right) \right). \label{dilatonpre2}
\end{align}


Expanding  near $Y_c \to \infty$ and keeping only the leading and subleading terms, the first term vanishes and $\operatorname{erf}\left(\frac{Y_c}{2 \sqrt{\Omega}}\right) \rightarrow 1$ and thus we obtain

\begin{align}
  Z_{\text{wall}} 
  &=-\frac{1}{2}\times 2  Z_{\text{nz}} \int d^{D-1} Y \  \partial_D \phi\left(Y^i, 0\right)    \left( \frac{\apr Y_c^2}{4} + \frac{\apr \Omega}{2} \right).
\end{align}
There is a factor of $e^{-2\Phi_0}$ sitting inside $Z_{\text{nz}}$. Extracting it gives $Z_{\text{nz}} = \tilde{Z}_{\text{nz}} e^{-2\Phi_0}$, where $\tilde{Z}_{\text{nz}}$ does not depend on the dilaton

\begin{align}
  Z_{\text{wall}} 
  &=-\frac{1}{2}\times 2  \tilde{Z}_{\text{nz}} \int d^{D-1} Y \ e^{-2\Phi_0} \partial_D \phi\left(Y^i, 0\right) \left( \frac{\apr Y_c^2}{4} + \frac{\apr \Omega}{2} \right)  .
\end{align}
We can express the derivative of $\phi$ as the derivative of $\Phi$ (as $\Phi_0$ is just a constant) and also replace $e^{-2\Phi_0}$ with $e^{-2\Phi}$ (as the difference will just be higher order) to finally get
\begin{align}
  Z_{\text{wall}} 
  &=-\frac{1}{2}\times 2  \tilde{Z}_{\text{nz}} \int d^{D-1} Y \ e^{-2\Phi} \partial_D \Phi \left(Y^i, 0\right)    \left( \frac{\apr Y_c^2}{4} + \frac{\apr \Omega}{2} \right) .
  \end{align}
  
Using that $\Omega = -\ln \eps$, $Z_{\text{wall}}$ takes the final form
\begin{align}
  Z_{\text{wall}} 
  &=-\frac{1}{2}\times 2  \tilde{Z}_{\text{nz}}\, \apr\int d^{D-1} Y \ e^{-2\Phi} \partial_D \Phi\left(Y^i, 0\right)     \left( \frac{ Y_c^2}{4} + \frac{- \ln \eps}{2} \right).
\end{align}

The above form should be familiar. It has a structure similar to that of the bulk sphere partition function: there is an overall log divergence multiplying the boundary dilaton sphere effective action.\footnote{Although we emphasize that in this work, we don't have an independent NLSM derivation of $\partial_D \Phi\left(Y^i, 0\right)$ in terms of the dilaton beta function.} 

As in the bulk off-shell action, we apply Tseytlin's prescription $\FTP$ to eliminate overall the log divergence and the divergent $Y_c^2$ term to obtain the classical off-shell boundary action for the dilaton in half-space
\begin{equation}\label{eq:I_wall}
I_{\text{wall}} = -\FTP Z_{\text{wall}} 
=-\frac{\apr}{2}\tilde{Z}_{\text{nz}}  \int d^{D-1} Y \ e^{-2\Phi} \partial_D \Phi\left(Y^i, 0\right) = +\frac{\apr}{2}\tilde{Z}_{\text{nz}}  \int d^{D-1} Y \ e^{-2\Phi} \partial_n \Phi\left(Y^i, 0\right) ,
\end{equation}
where $\partial_n$ is the normal derivative in the direction of the outward-pointing normal $n^\mu$ i.e. $\partial_n=-\partial_D$.


Now, we compute $Z_{\text{far}}$ to obtain the bulk contribution to $Z_{\text{HS}}$ 
\begin{align}
Z_{\text{far}}&=\frac{1}{2}Z_{\text{nz}} \int_{\frac{\abs{Y^D}}{\sqrt{\apr}}\gg 1} d^D Y  \left(-\frac{1}{4 \pi } \int d^2 z \sqrt{g}   R^{(2)} \left\langle \tilde{\phi}(Y^{\mu}+\sqrt{\apr}\eta^{\mu}(z))\right\rangle\right) \\
&=\frac{1}{2}Z_{\text{nz}} \int_{\frac{\abs{Y^D}}{\sqrt{\apr}}\gg 1} d^D Y  \left(-\frac{1}{4 \pi } \int d^2 z \sqrt{g}   R^{(2)} \left(\tilde{\phi}\left(Y^{\mu}\right) + \frac{\apr}{2} \partial_\mu \partial_\nu \tilde{\phi}\left(Y^{\mu}\right) \left\langle \eta^\mu(z) \eta^\nu (z) \right\rangle + O(\alpha^{\prime 3/2})\right)\right).
\end{align}
Keeping only the relevant term (with a log $\eps^{-1}$ coefficient) in $Z_{\text{far}}$ is
\begin{align}\label{eq:Z_far}
Z_{\text{far}}
&=\frac{1}{2}Z_{\text{nz}} \int_{\frac{\abs{Y^D}}{\sqrt{\apr}}\gg 1} d^D Y  \left(-\frac{1}{4 \pi } \int d^2 z \sqrt{g}   R^{(2)}  \frac{\apr}{2} \partial_\mu \partial_\nu \tilde{\phi}\left(Y^{\mu}\right)  \delta^{\mu \nu}  \left(- \frac{1}{2}\log \eps  + O(\eps^2)\right) \right).
\end{align}

Applying Tseytlin's prescription to $Z_{\text{far}}$ gives $I_{\text{bulk}}$
\begin{align}
I_{\text{bulk}} := I_{\text{far}} &= -\FTP Z_{\text{far}} 
\\
&= - \frac{1}{4}Z_{\text{nz}} \frac{\apr}{2} \int_{\frac{\abs{Y^D}}{\sqrt{\apr}}\gg 1} d^D Y  \left(\frac{1}{4 \pi } \int d^2 z \sqrt{g}   R^{(2)}   \partial^2 \tilde{\phi}\left(Y^{\mu}\right)\right)\\
&= - \frac{2}{4}Z_{\text{nz}} \frac{\apr}{2} \int_{\frac{Y^D}{\sqrt{\apr}}\gg 1} d^D Y  \left(\frac{1}{4 \pi } \int d^2 z \sqrt{g}   R^{(2)}   \partial^2 \tilde{\phi}\left(Y^{\mu}\right)\right)\\
&= - \frac{2}{4}Z_{\text{nz}} \frac{\apr}{2} \int_{\frac{Y^D}{\sqrt{\apr}}\gg 1} d^D Y  \left(\frac{8 \pi}{4 \pi }    \partial^2 \tilde{\phi}\left(Y^{\mu}\right)\right)\\
&=  -\frac{1}{2}\tilde{Z}_{\text{nz}} \apr \int_{\frac{Y^D}{\sqrt{\apr}}\gg 1} d^D Y  e^{-2\Phi}  \partial^2\Phi \label{eq:FreeLapPhi}
\end{align}

This free action for the dilaton \eqref{eq:FreeLapPhi} does \textit{not} have a well-defined variational principle since upon variation, it gives
\begin{align}
\delta I_{\text {bulk }}&=  \alpha^{\prime} \tilde{Z}_{\text{nz}} \int_{\frac{Y^D}{\sqrt{\apr}}\gg 1} d^D Y\left[e^{-2 \Phi} \partial^2 \Phi+\partial_\mu\left(e^{-2 \Phi} \partial^\mu \Phi\right)\right] \delta \Phi \nonumber \\
&- \frac{1}{2}\apr \tilde{Z}_{\text{nz}} \int_{\partial \mathcal{M}} d^{D-1} Y  e^{-2\Phi} \, \delta(\partial_n \Phi)  - \alpha^{\prime} \tilde{Z}_{\text{nz}} \int_{\partial \mathcal{M}} d^{D-1} Y e^{-2 \Phi} \partial_n \Phi \, \delta \Phi . \label{eq:deltaI_bulk}
\end{align}

The first term in \eqref{eq:deltaI_bulk} is the bulk dilaton equation of motion while the second and third are boundary terms. The third term vanishes for Dirichlet boundary conditions. The second term is what spoils the variational principle for Dirichlet boundary condition. It is the GHY-like term of the free two-derivative action \eqref{eq:FreeLapPhi} which must be accounted for, i.e. by adding an \textit{additional} boundary term whose variation exactly cancels with it. This additional boundary term is $I_{\text{wall}}$ in \eqref{eq:I_wall}. Let's see how this works.

The total classical off-shell string action for the dilaton in half-space is the \textit{sum} of bulk and boundary contributions \eqref{eq:I_wall}
\begin{align}
I_{\text{sphere}}:= I_{\text{HS}}&=I_{\text{bulk}}+I_{\text{wall}} \nonumber\\
&=-\frac{1}{2}\tilde{Z}_{\text{nz}} \apr \int_{\frac{Y^D}{\sqrt{\apr}}\gg 1} d^D Y  e^{-2\Phi}    \partial^2 \Phi  + \frac{1}{2}\apr\tilde{Z}_{\text{nz}}  \int d^{D-1} Y \ e^{-2\Phi} \partial_n \Phi\left(Y^i, 0\right) \\
&=- \tilde{Z}_{\text{nz}} \apr \int_{\frac{Y^D}{\sqrt{\apr}}\gg 1} d^D Y  e^{-2\Phi}    \partial_\mu \Phi\partial^\mu \Phi,
\label{eq:I_HS}
\end{align}
where integration by parts is used to get the last line.

Varying $I_{\text{sphere}}$ gives
\begin{align}
\delta I_{\text {sphere}} =&\, \delta I_{\text {bulk }} + \delta I_{\text {wall}} \nonumber  \\
&=  \alpha^{\prime} \tilde{Z}_{\text{nz}} \int_{\frac{Y^D}{\sqrt{\apr}}\gg 1} d^D Y\left[e^{-2 \Phi} \partial^2 \Phi+\partial_\mu\left(e^{-2 \Phi} \partial^\mu \Phi\right)\right] \delta \Phi \\
&- \frac{1}{2}\apr \tilde{Z}_{\text{nz}} \int_{\partial \mathcal{M}} d^{D-1} Y  e^{-2\Phi} \, \delta(\partial_n \Phi) + \frac{1}{2}\apr \tilde{Z}_{\text{nz}} \int_{\partial \mathcal{M}} d^{D-1} Y  e^{-2\Phi} \, \delta(\partial_n \Phi)  \\
&-2 \alpha^{\prime} \tilde{Z}_{\text{nz}} \int_{\partial \mathcal{M}} d^{D-1} Ye^{-2 \Phi} \partial_n \Phi \, \delta \Phi \\
&= \alpha^{\prime} \tilde{Z}_{\text{nz}} \int_{\frac{Y^D}{\sqrt{\apr}}\gg 1} d^D Y\left[e^{-2 \Phi} \partial^2 \Phi+\partial_\mu\left(e^{-2 \Phi} \partial^\mu \Phi\right)\right] \delta \Phi -2 \alpha^{\prime} \tilde{Z}_{\text{nz}} \int_{\partial \mathcal{M}} d^{D-1} Ye^{-2 \Phi} \partial_n \Phi \, \delta \Phi.
\end{align}
In the second line, we notice that the variation $\delta I_{\text{wall}}$ exactly cancels its counterpart in $\delta I_{\text{bulk}}$ i.e. $ e^{-2\Phi} \, \delta(\partial_n \Phi)$.  Therefore, $I_{\text {sphere}}$ has a well-defined variational principle only with the addition of $I_{\text{wall}}$ as a boundary term. 
\\~\\
Having derived this term from the worldsheet is the main result of this paper.

\subsection{The classical on-shell action}\label{sec:On-shell_Action}
Let us now impose the dilaton equation of motion in $I_{\text {bulk}}$ to obtain the on-shell action. To do that, we first integrate by parts \eqref{eq:FreeLapPhi}  to obtain
\begin{align}
I_{\text{bulk}} &=  -\frac{1}{2}\tilde{Z}_{\text{nz}} \apr \int_{\frac{Y^D}{\sqrt{\apr}}\gg 1} d^D Y  e^{-2\Phi}    \partial^2 \Phi  \\
&= -\frac{1}{2}\tilde{Z}_{\text{nz}} \apr \int_{\frac{Y^D}{\sqrt{\apr}}\gg 1} d^D Y\, \partial_\mu\left( e^{-2\Phi}  \partial^\mu \Phi\right) -\tilde{Z}_{\text{nz}} \apr \int_{\frac{Y^D}{\sqrt{\apr}}\gg 1} d^D Y   e^{-2\Phi}  (\partial \Phi)^2\\
&= -\frac{1}{2}\tilde{Z}_{\text{nz}} \apr \int d^{D-1} Y e^{-2\Phi} \partial_n\Phi -\tilde{Z}_{\text{nz}} \apr \int_{\frac{Y^D}{\sqrt{\apr}}\gg 1} d^D Y e^{-2\Phi}  (\partial \Phi)^2 \,.\label{eq:I_far}
\end{align}
Imposing the dilaton equation of motion in $I_{\text{bulk}}$ 
\begin{equation}
e^{-2\Phi}  (\partial \Phi)^2 = - \partial_\mu\left( e^{-2\Phi} \partial^\mu \Phi\right)
\end{equation}
leaves us with the following bulk on-shell action
\begin{equation}\label{eq:I_bulk_onshell}
I_{\text{bulk}} = +\frac{1}{2}\tilde{Z}_{\text{nz}} \apr \int d^{D-1} Y e^{-2\Phi} \partial_n\Phi
\end{equation}

The classical on-shell action for the dilaton in half-space is therefore the sum of the two boundary contributions \eqref{eq:I_bulk_onshell} and \eqref{eq:I_wall}

\begin{align} \label{eq:I_onshell}
I_{\text{sphere}}:=I_{\text{HS}}&=I_{\text{bulk}}+I_{\text{wall}} =  \, \apr \tilde{Z}_{\text{nz}} \int d^{D-1} Y e^{-2\Phi} \partial_n\Phi \,.
\end{align}
Therefore, we observe that the two boundary terms in $I_{\text{sphere}}$ add up to give the on-shell classical action.


\section{Discussion} \label{sec:discussion}


In this section, we begin by making the following observation: the expectation value of $\abs{\eta^D(z_0)}^N$ \eqref{eq:n-point-corr}
\begin{align}\label{eq:EtaMom}
 \left\langle \abs{\eta^D(z_0)}^N \right\rangle
 &= \frac{1}{\sqrt{\pi}}  \Gamma\left(\frac{1 + N}{2}\right)  \Omega^{N/2}\,,
\end{align}
corresponds to the $N$-th moment of the half-normal distribution for the random variable $|X|$. We will then discuss the relationship between the two quantities and make observations that connect them to the theory of reflected Brownian motion (RBM) and one-dimensional random walks in spaces with a reflecting barrier. We begin by defining the half-normal distribution and its key properties.


Let $X_{t}$ be a normally distributed random variable $N(0,t^2)$ with mean 0  and standard deviation $t$. The half-normal distribution of {$Y_t= |X_{t}|$} is a one-parameter scale family of distributions with scale parameter $t \in [0, \infty)$. It is a special case of the folded normal distribution where $X_{t}$ has nonzero mean.

The probability density function (PDF) $f(y;t)$ of $Y$ with scale parameter $t$ is given by

\begin{equation}\label{eq:HNDPD}
f(y;t) = \frac{1}{t} \sqrt{\frac{2}{\pi}} \exp\left(-\frac{y^2}{2 t^2}\right), \quad y \in [0, \infty)\,,
\end{equation}
while the cumulative distribution function (CDF) $F(y;t)$ is
\begin{equation}
F(y;t) = \int_0^y \frac{1}{t} \sqrt{\frac{2}{\pi}} \exp\left(-\frac{x^2}{2 t^2}\right) \, dx \,.
\end{equation}


For \textit{any} (even or odd) $N \geq 0$, the \textit{absolute} central moments are given by \cite{HNDMathWorks}\footnote{We assume the expectation value (first central moment) of $X$ is zero. Also, we are using $\langle X^N \rangle$ to denote the moments of $X$ rather than $\operatorname{E}[X^N]$.} 
\begin{align}\label{eq:AbsMom}
\left\langle Y_{t}^N\right\rangle = {t^N}\left\langle|X_{t}|^N\right\rangle
&= t^N \cdot \frac{2^{N/2}\Gamma\left(\frac{N+1} 2 \right)}{\sqrt\pi},
\end{align}
with variance $\operatorname{var}(Y_{t})=t^2\left(1-\frac{2}{\pi}\right)$.
For $N=1$, $\left\langle|X|\right\rangle =  \frac{t \sqrt{2}}{\sqrt{\pi}}$ is the mean of the half-normal distribution. 

Comparing \eqref{eq:EtaMom} with \eqref{eq:AbsMom}, we observe that they are identical 
\begin{equation}
\left\langle \abs{\eta^D(z_0)}^N \right\rangle = \left\langle \abs{X}_{t}^N\right\rangle \,,
\end{equation}
if we make the identification
\begin{equation}
\Omega^{N/2} = t^N 2^{N/2},  
\end{equation}
which means $\Omega$ must be
\begin{equation}
\Omega = 2 t^2.
\end{equation}
Using $\Omega = \log \epsilon^{-1}$ \eqref{eq:Omega_ln_eps}, the scale parameter $t$ of the distribution is identified with the worldsheet UV cutoff $\eps$ in the following way
\begin{equation}
t\sqrt{2}  = \sqrt{-\log \eps} \,.
\end{equation} 

In particular, the one-point function of the string nonzero mode $\abs{\eta^D(z_0)}$ corresponds to the first moment $(N=1)$ of the random variable $\abs{X_{t}}$
\begin{equation}\label{eq:HND_iden}
\left\langle \abs{\eta^D(z_0)}\right\rangle =\left\langle \abs{X_{t}}\right\rangle  = \frac{t \sqrt{2}}{\sqrt{\pi}}.
\end{equation}

This observation gives a probabilistic interpretation of $I_{\text{HS}}$ and connects it to the vast and interesting literature on the theory of reflected Brownian motion  \cite{morters2010brownian,karatzas2014brownian}. The half-normal distribution naturally appears in the theory of stochastic (Wiener) processes of $\left\langle \abs{X_{t}}\right\rangle$ in spaces with reflecting boundaries. Specifically, it arises as the probability distribution for the running maximum \cite{morters2010brownian,karatzas2014brownian,borodin2015handbook,grebenkov2019probability} in an RBM. See for \cite{random} for less technical definitions of RBM and its key properties. In \cite{banderier2018local,wallner2020half}, it is also shown that the half-normal distribution emerges as the limit law for the \textit{local time} at $x=0$ in directed lattice path on $\mathbb{Z}$ with no drift. The local time at zero in this type of random walk is the sum the number of returns to zero and number of $x$-axis crossings. 

\begin{figure*}
\centering
\begin{subfigure}{.5\textwidth}
  \centering
  \includegraphics[width=.85\linewidth]{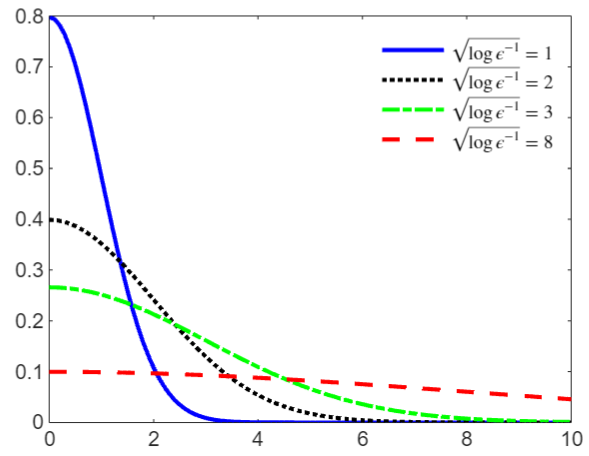}
  \caption{The half-normal PDF $f(y;t)$}
  \label{fig:PDFHND}
\end{subfigure}%
\begin{subfigure}{.6\textwidth}
  \centering
  \includegraphics[width=.72\linewidth]{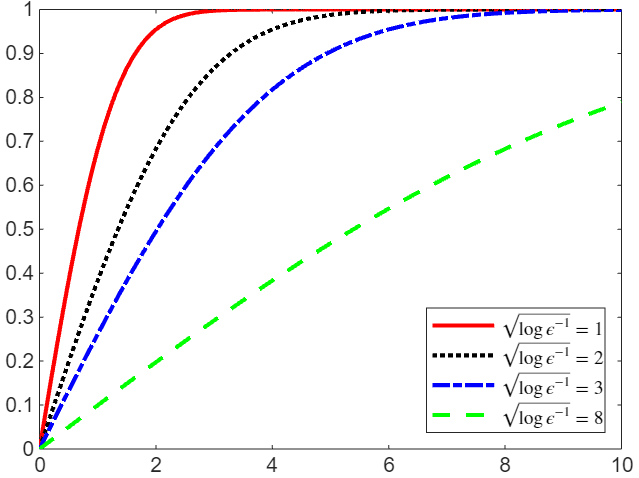}
  \caption{The half-normal CDF $F(y;t)$}
  \label{fig:CDFHND}
\end{subfigure}
\caption{The half-normal probability density $f(y;t)$ and cumulative distribution $F(y;t)$ functions for different scale parameters $t = \frac{\sqrt{\log \eps^{-1}}}{\sqrt{2}}$. The effect of changing the value of the UV cutoff $\eps$ is to change the overall size of the distribution. A larger value of $t$ (or equivalently, a smaller value of $\eps$) brings the probability density on the left closer to a delta function  and the distribution on the right closer to a step function $\theta(Y)$. Figures adapted from \url{https://www.mathworks.com/help/stats/half-normal-distribution.html}.}
\label{fig:HND}
\end{figure*}

We would like to comment on the relationship in \eqref{eq:HND_iden} between the UV cutoff $\eps^{-1}$ and the parameter $t$. From \eqref{eq:HND_iden},  $\eps^{-1}$ appears \textit{in target space} as a dial that controls the \textit{overall} scale $t$ of the half-normal distribution. Each value of $\eps^{-1}$ describes a worldsheet theory at a particular renormalization group (RG) scale. Fig. \ref{fig:HND} shows the PDF and CDF of the half-normal probability distribution with different values of the scale parameter $t = \sqrt{\frac{-\log \eps}{2}}$. The plots may be interpreted as the RG flow from the UV with a larger value of $\eps^{-1}$ (or smaller value of $\log \eps^{-1}$) to the IR with a smaller value of $\eps^{-1}\rightarrow 0$. In this sense, the RG flow renormalizes the length of the Euclidean Schwinger propagator or equivalently, the Schwinger time. As explained in \cite{Ahmadain:2022tew}, the $\log \eps^{-1} \rightarrow 0$ limit corresponds to the S-matrix regime whereas when $\eps \sim 1$, the worldsheet theory is in the extreme UV limit.

\subsection{The probabilistic nature of the $\left\langle\abs{Y^D+\sqrt{\apr}\eta^D(z)}\right\rangle$ }\label{sec:Prob-One-Point-Func}

Reflected Brownian motion is the formal mathematical framework used to describe this probabilistic interpretation of the stochastic string dynamics in target spacetime. We argue in this section that $\abs{\eta^D(z)}$ has the same probability distribution of the boundary local time (defined below)
\begin{equation}
\left\langle\abs{\eta^D(z)}\right\rangle = \EV{L^x_t},
\end{equation}
thereby giving a probabilistic description of the string dynamics in the target space. 

The discussion in this section is speculative in nature and, as such, is not rigorous. However, we will make use of of well-known concepts and notions in the theory of RBM that may be unfamiliar to the typical reader of this paper. For brevity, we will not provide the proper mathematical definitions here but rather give an extensive list of references for the interested reader. In particular, we will use the Skorokhod construction \cite{skorokhod1961stochastic,skorokhod1962stochastic}, Tanakas's formula \cite{tanaka1979stochastic} and the important notion of local boundary time \cite{levy1965processus,it1965diffusion}.

In worldline path integrals, the exponential of the boundary local time gives the total probability of particle paths reflecting from the wall \cite{Clark1980half}. The boundary local time measures the amount of time a Brownian motion $\boldsymbol{X}$ spends at level $x$. Informally, it is defined as 
\begin{equation}\label{eq:LBT}
L^x_t=\int_0^t \delta_x\left(X_s\right) d s.
\end{equation}

A more formal definition of $L_t$ is given in terms of the limit of the occupation time $T_t$ of the Brownian motion $\textbf{X}_t$ in a region $\varepsilon$ near the boundary $\partial\mathcal{M}_{\varepsilon} = \{X_t \in \mathcal{M} : |x-\partial \mathcal{M}| < \varepsilon\}$

\begin{equation}\label{eq:Reg_LBT}
L_t=\lim _{\varepsilon \rightarrow 0} \frac{1}{2\varepsilon} \int_0^t ds \,\mathbb{1}_{\partial \mathcal{M}_\varepsilon}\left({X}_s\right),
\end{equation}
where $\mathbb{1}_{\partial \mathcal{M}_{\varepsilon}}(X_s)$ is the indicator function of the boundary (i.e., $\mathbb{I}_{\partial \mathcal{M}_{\varepsilon}}(X_t)=1$ if $X_t \in \partial \mathcal{M}_{\varepsilon}$, and 0 otherwise). 
With this definition, the local boundary time $L_t$ measures the amount of time $X_{t}$ spends in a region $\varepsilon$ \textit{near} the boundary $\partial \mathcal{M}$, and thus increases \textit{only} when ${X}_s \in \partial \mathcal{M}_{\varepsilon}$.

Let our bounded region in $\mathbb{R}^D$ be the half-space $\mathbb{R}_+ \times \mathbb{R}^{D-1}$. In this case, it is known that Tanaka's formula \cite{tanaka1979stochastic} solves the Skorokhod equation \cite{skorokhod1961stochastic, skorokhod1962stochastic} for the reflected process $\left|X_t\right|$.\footnote{Specifically, for the process $\left|X_t\right|=\left|Y^D\right|+\int_0^t \operatorname{sgn}\left(X_s\right) d X_s+L_t^0$ where  $\operatorname{sgn}\left(X_s\right)$ is the sign function of $X_s$ and $L_t^0$ is the local time of $X_t$ at 0 up to time $t$. Here, $L_t^0$ is the non-decreasing \textit{regulator} process.} 
Using the reflection principle, it is also known that the following three processes are equal in distribution law \cite{mckean1969stochastic}
\begin{equation}\label{eq:ProbLaw}
L^0_t(X) \overset{d}{=} S_t \,,
\end{equation}
where in the notation of \cite{björk2015pedestrians}, $\overset{d}{=}$ denotes equality in distribution $S_t$ = $\sup_{s \leq t}$ is the running maximum of the Brownian motion.

Let the string's motion start at $Y^D = 0$, right at the wall. Taking the spatial expectation value (the first moment) of \eqref{eq:ProbLaw} in half-space, then get at a given fixed time $t$, we get
\begin{equation}\label{eq:ProbLaw_One-point}
\EV{S_t} = \EV{L^0_t(X)}.
\end{equation}
But $S_t$ is known to be half-normally distributed in half-space \cite{takacs1995local,random}. Therefore, using \eqref{eq:HND_iden} and \eqref{eq:ProbLaw_One-point}, we may conclude that $\abs{\eta^D(z_0)}$ can be interpreted as the boundary local time $L^0_t(X)$ in half-space with the same probability distribution
\begin{equation}\label{eq:eta_LBT_iden}
\left \langle\abs{\eta^D(z_0)}\right\rangle \overset{d}{=}\EV{L^0_t(X)} = \frac{\sqrt{\Omega}}{\sqrt{\pi}} = \frac{\sqrt{- \log \eps}}{\sqrt{\pi}}\,.
\end{equation}


To recap, we have argued above that the string nonzero mode $\abs{\eta^D(z_0)}$  has the same spacetime dynamics of a standard Brownian motion (with zero mean) away from the wall whereas near the boundary, $\abs{\eta(z)}$, it is equivalent \textit{in distribution law} to boundary local time $L^0_t(X)$. This provides a justification for splitting $Z_{\text{HS}}$ in section \ref{sec:BoundaryAction} into $Z_{\text{near}}$ and $Z_{\text{far}}$. In this distributional sense, the one-point function $\left\langle \left|Y^D + \sqrt{\apr} \eta^D(z_0)\right|\right\rangle$ measures the average length (or time) of the string near the wall at $X_D=0$ taking into account the total number of bounces off of it. Therefore, generally speaking, in spaces with reflecting barriers, for different types of reflecting boundary conditions, we expect string boundary physics to have a probabilistic interpretation\footnote{For example, a sticky motion is a boundary condition where the string can stick to the boundary for some time before getting reflected.}.

We would like to make one comment on the one-point function $u(Y^D) \coloneqq \left\langle \left|Y^D + \sqrt{\apr} \eta^D(z_0)\right|\right\rangle$. It  turns out $u(Y^D)$ is the solution to the one-dimensional heat equation with a non-homogeneous source term
\begin{equation}
\frac{d^2 u}{d\left(Y^D\right)^2}=\frac{2}{\sqrt{\alpha^{\prime} \pi \Omega}} e^{-\frac{\left(Y^D\right)^2}{\alpha^{\prime} \Omega}} .
\end{equation}
where the explicit evaluation of $u(Y^D)$ gives
\begin{align}\label{eq:u(Y)_sol}
u(Y^D) & =\sqrt{\Omega} \frac{\sqrt{\alpha^{\prime}}}{\sqrt{\pi}} e^{-\frac{\left(Y^D\right)^2}{\alpha^{\prime} \Omega}}+Y^D \operatorname{erf}\left(\frac{Y^D}{\sqrt{\alpha^{\prime} \Omega}}\right) \\
& =\int_0^{Y^D} d Y^D \operatorname{erf}\left(\frac{Y^D}{\sqrt{\alpha^{\prime} \Omega}}\right)+ C .
\end{align}
With the following boundary conditions
\begin{equation}
u'(0) = \operatorname{erf}\left(\frac{0}{\sqrt{\alpha' \Omega}}\right) = 0, \quad u'(\infty) = \operatorname{erf}\left(\frac{\infty}{\sqrt{\alpha' \Omega}}\right) = 1, \quad u(0) = \sqrt{\Omega} \frac{\sqrt{\alpha^{\prime}}}{\sqrt{\pi}} = \sqrt{\alpha^{\prime}}\left\langle\left|\eta^D\left(z_0\right)\right|\right\rangle,
\end{equation}
the constant is $C = u(0)$, which is the value of $u(Y^D)$ at $Y^D=0$, as found in appendix \ref{sec:coincorrfunc}. Therefore, we see that $u(Y^D)$ is the general solution of an equation with a Gaussian source. The Gaussian is centered at $Y^D = 0$ with a variance proportional to $\alpha'\Omega$. As $\Omega \to 0$ ($ -\ln \eps \to 0$), the source becomes increasingly peaked and narrow around zero approaching a Dirac delta function $\delta(Y^D)$. This is consistent with the RG flow picture explained earlier. See Fig. \ref{fig:HND}.

We end this discussion by making one final observation. As mentioned in the beginning of this section, the half-normal distribution emerges as the \textit{limit} law for the \textit{local time} at $x=0$ in directed lattice paths (random walks) on $\mathbb{Z}$ with no drift in the limit of large number of steps \cite{banderier2018local,wallner2020half}. Specifically, 
\begin{equation}\label{eq:RW_return}
\textrm{local time at zero = the number of returns to zero} + \text{number of } x\text{-axis crossings .} 
\end{equation}

In light of the argument presented earlier \eqref{eq:eta_LBT_iden}, is it possible to interpret the motion of the string's nonzero mode $\abs{\eta^D(z)}$ as a random walk with an average number of returns to the wall as defined in \eqref{eq:RW_return}? It would be great to further study the relationship between strings and random walks (for example, see section 6.4 in \cite{Ahmadain:2022gfw}), but we leave this for future work.


\subsection{Outlook and Future Directions}
A generalization of the current work to $S^1/\mathbb{Z}_2$ and curved target spaces with boundaries is an interesting direction. The generalization to codimension-2 boundaries, especially conical geometries are all interesting future directions. A worldsheet understanding of how to impose Dirichlet boundary conditions on propagating fields in target spacetime is required to derive the GHY boundary term in the on-shell classical action. This, in turn, is important for deriving the classical black hole entropy \cite{GH}. 

A probabilistic understanding of the boundary local time in conical geometries may give us some clues that might help us unravel the nature of the boundary conditions required for a worldsheet derivation of the classical black hole entropy $\frac{A}{4 G_N}$ and the Ryu-Takayanagi formula \cite{RTPhysRevLett.96.181602} as presented in \cite{LM2013}. In particular, figuring out the boundary local time in conical geometries near the tip of the cone may guide us to understand the boundary condition in the NLSM limit of the winding condensate \cite{Jafferis-ER=EPR:2021,Halder:2023adw}.

In \cite{Ahmadain:2024uyo}, the method of images is used to derive the boundary term of the Einstein-$\Gamma^2$ action \cite{Einstein:1916cd} from the spherical worldsheet to first order in $\alpha^{\prime}$ and to linear order in the metric perturbation around flat half-space. The $\Gamma^2$ action has a boundary term that consists of the Gibbons-Hawking-York action \cite{York:1972sj,Gibbons:1976ue} (along with two additional terms). With the GHY term, the total sphere effective action has a well-posed variational principle for Dirichlet boundary conditions.


Another very interesting direction involves understanding the behavior of strings at a \textit{finite non-asymptotic} Dirichlet boundary on a bulk Cauchy slice in $\operatorname{AdS}$ \cite{Zamolodchikov:2004ce,Smirnov:2016lqw,Cavaglia:2016oda,McGough:2016lol,Araujo-Regado:2022gvw} for T$\bar{\text{T}}$-deformed CFTs. The related question of placing a finite timelike boundary in an $\operatorname{AdS}$ black hole and in a $\operatorname{dS}$ static patch requires an understanding of the behavior of strings near totally absorbing walls with a Dirichlet boundary condition \cite{silverstein2023black,Batra:2024kjl}.\footnote{We thank Eva Silverstein and Aron Wall for a discussion of this point.} It would be very interesting if the work in this paper can be generalized to address this question.


\section*{Acknowledgements}
We are grateful to Aron Wall, Eva Silverstein, Daniel Jeffries, Alex Frenkel, Raghu Mahajan, Prahar Mitra, Arkadi Tseytlin, Edward Witten, Jose Sa, Xi Yin, Gabriel Wong, Yiming Chen, Douglas Stanford, Ronak Soni, David Kolchmeyer, and Juan Maldacena for insightful discussions and comments. We thank Shoaib Akhtar for critically reviewing version 1 of this paper. AA thanks Minjae Cho, Harold Erbin, Lorenz Eberhardt, and Daniel Thompson for interesting discussions. We are especially grateful to Aron Wall for very insightful discussions during the course of this project, reviewing parts of the manuscript and providing valuable feedback and comments. We are also grateful to Eva Silverstein for reviewing parts of the manuscript and giving valuable comments. AA is grateful to Edward Witten and Juan Maldacena for the opportunity to give a talk about this work while it was still ongoing. AA is supported by The Royal Society and by STFC Consolidated Grant No. ST/X000648/1.
\paragraph{}
{\footnotesize {\bf Open Access Statement} - For the purpose of open access, the authors have applied a Creative Commons Attribution (CC BY) licence to any Author Accepted Manuscript version arising.}

\footnotesize{\textbf{Data access statement}: no new data were generated for this work.}

\begin{appendices}
\section{The computation of $\left\langle\abs{Y^D+\sqrt{\apr}\eta^D(z)}\right\rangle$}\label{sec:coincorrfunc}

In this section we show how to compute the one-point function $\left\langle\abs{Y^D+\sqrt{\apr}\eta^D(z)}\right\rangle$. Let us in fact be more general and compute $\left\langle f\left(\eta^D(z)\right)\right\rangle$ for an arbitrary function $f\left(\eta^D(z)\right)$.

The first step is to expand $\eta^D(z)$ in spherical harmonics
\begin{align}
  \eta^D(z) = \sum_{l=1}^\infty \sum_{m=-l}^{m=l}   x_{lm} Y_{lm}(z),
\end{align}
where $Y_{lm}(z)$ are the real spherical hamonics 
\begin{equation}
Y_{\ell m} = \begin{cases} 
\left(-1\right)^{m} \sqrt{2} \sqrt{\dfrac{2\ell +1}{4\pi} \dfrac{(\ell -|m|)!}{(\ell +|m|)!}} P_{\ell}^{|m|}(\cos \theta) \sin(|m|\varphi) & \text{if } m<0 \\
\sqrt{\dfrac{2\ell +1}{4\pi}} P_{\ell}^{m}(\cos \theta) & \text{if } m=0 \\
\left(-1\right)^{m} \sqrt{2} \sqrt{\dfrac{2\ell +1}{4\pi} \dfrac{(\ell -m)!}{(\ell +m)!}} P_{\ell}^{m}(\cos \theta) \cos(m\varphi) & \text{if } m>0 \,.
\end{cases}
\end{equation}
They satisfy the Laplace equation in spherical coordinates 
\begin{equation}
\nabla^2 Y_{lm}(z) = -l(l+1)Y_{lm}(z)\,,
\end{equation}
and are orthogonal
\begin{equation}
\int_0^{2\pi} d\phi \int_0^\pi d\theta \  \sin{\theta}  \  Y_{lm}(\theta,\phi)  Y_{l'm'}(\theta,\phi) = \delta_{ll'} \delta_{mm'}\,,
\end{equation}
where $x_{lm}$ are the expansion coefficients, and the sum runs from $l=1$ to $\infty$, not including the zero mode. Now we have
\begin{align}
\left\langle f\left(\eta^D(z)\right)\right\rangle &=\frac{1}{Z_{\text{nz}}} \int [D\sqrt{\apr}\eta] e^{-I_{\text{p}}[\eta,g]} f\left(\eta^D(z)\right)\\
&=\frac{1}{Z^{(D)}_{\text{nz}}} \int [D\sqrt{\apr}\eta^D] e^{-I^{(D)}_{\text{p}}[\eta^D,g]} f\left(\eta^D(z)\right),
\end{align}
where in the second line, we used the fact that the Polyakov action factorises into a sum of two terms: a term containing $\eta^D$ and another term containing $\eta^i$. Now we multiply by $1$ and express $1$ as a Dirac delta function
\begin{align}
\left\langle f\left(\eta^D(z)\right)\right\rangle
&=\frac{1}{Z^{(D)}_{\text{nz}}} \int [D\sqrt{\apr}\eta^D] e^{-I^{(D)}_{\text{p}}[\eta^D,g]} f\left(\eta^D(z)\right) 1\\
&=\frac{1}{Z^{(D)}_{\text{nz}}} \int [D\sqrt{\apr}\eta^D] e^{-I^{(D)}_{\text{p}}[\eta^D,g]} f\left(\eta^D(z)\right) \int_{-\infty}^{\infty} d\xi \ \delta\left(\xi-\eta^D(z)\right)\\
&=\frac{1}{Z^{(D)}_{\text{nz}}} \int_{-\infty}^{\infty} d\xi  \int [D\sqrt{\apr}\eta^D] e^{-I^{(D)}_{\text{p}}[\eta^D,g]} f\left(\sum_{l=1}^\infty \sum_{m=-l}^{m=l}   x_{lm} Y_{lm}(z_0)\right) \ \delta\left(\xi-\sum_{l=1}^\infty \sum_{m=-l}^{m=l}   x_{lm} Y_{lm}(z)\right),
\end{align}
where the measure in terms of the expansion coefficients is now
\begin{align}
  [D\sqrt{\apr}\eta^D] =   \prod_{l=1}^\infty \prod_{m=-l}^{m=l} \sqrt{\apr} dx_{lm}\,.
\end{align}

The Polyakov action of the $\eta^D(z_0)$ is given by (using the round sphere worldsheet metric)
\begin{align}
   I^{(D)}_{\text{p}}[\eta^D,g]&= \frac{1}{4 \pi \apr} \int d^2 z \sqrt{g}\left(\apr g^{ab}\partial_a \eta^D \partial_b \eta^D    \right) + 2 \Phi_0 \\
   &= -\frac{1}{4 \pi \apr} \int d^2 z \sqrt{g}\left(\apr  \eta^D \nabla^2 \eta^D    \right) + 2 \Phi_0 \, .
\end{align}

The heat kernel-regularized $I^{(D)}_{\text{p}}[\eta^D,g]$ is given by 
\begin{align}
   I^{(D)}_{\text{p}}[\eta^D,g]
   &= -\frac{1}{4 \pi} \int d^2 z \sqrt{g}\left(  \eta^D e^{-\epsilon\nabla^2} \nabla^2 \eta^D    \right) + 2 \Phi_0 \\
   &= -\frac{1}{4 \pi}    \sum_{l=1}^\infty \sum_{m=-l}^{m=l}  (-l(l+1)) e^{\epsilon l(l+1)} x^2_{lm}     + 2 \Phi_0.
\end{align}

Thus, $\left\langle f\left(\eta^D(z)\right)\right\rangle$ is now given by
\begin{align}
\left\langle f\left(\eta^D(z)\right)\right\rangle
&=\frac{e^{-2\Phi_0}}{Z^{(D)}_{\text{nz}}} \int_{-\infty}^{\infty} d\xi  \int \left(\prod_{l=1}^\infty \prod_{m=-l}^{m=l} \sqrt{\apr} dx_{lm}\right) e^{-\frac{1}{4 \pi}    \sum_{l=1}^\infty \sum_{m=-l}^{m=l}  l(l+1) e^{\epsilon l(l+1)} x^2_{lm}} \\ &f\left(\sum_{l=1}^\infty \sum_{m=-l}^{m=l}   x_{lm} Y_{lm}(z_0)\right) \ \delta\left(\xi-\sum_{l=1}^\infty \sum_{m=-l}^{m=l}   x_{lm} Y_{lm}(z_0)\right).
\end{align}
Without loss of generality, we can take $z$ to be the north pole.\footnote{We could have proceeded without picking any point on the sphere at the expense of complicating the calculation bit.} This is because $\left\langle f\left(\eta^D(z)\right)\right\rangle$is independent of $z$ due to translation symmetry of the scalar field $\eta^D(z)$ on the worldsheet, which is \textit{not} broken even off-shell. Thus, we can just evaluate it at $z=z_0=0$
\begin{align}
\left\langle f\left(\eta^D(z_0)\right)\right\rangle
&=\frac{e^{-2\Phi_0}}{Z^{(D)}_{\text{nz}}} \int_{-\infty}^{\infty} d\xi  \int \left(\prod_{l=1}^\infty \prod_{m=-l}^{m=l} \sqrt{\apr} dx_{lm}\right) e^{-\frac{1}{4 \pi}    \sum_{l=1}^\infty \sum_{m=-l}^{m=l}  l(l+1) e^{\epsilon l(l+1)} x^2_{lm}} \\ &f\left(\sum_{l=1}^\infty \sum_{m=-l}^{m=l}   x_{lm} Y_{lm}(0)\right) \ \delta\left(\xi-\sum_{l=1}^\infty \sum_{m=-l}^{m=l}   x_{lm} Y_{lm}(0)\right).
\end{align}
At the north pole ($z_0=0$), the spherical hamonics vanish unless $m=0$, i.e. $Y_{lm}(0)=0$ when $m \neq 0$. So we get
\begin{align}
\left\langle f\left(\eta^D(z_0)\right)\right\rangle
&=\frac{e^{-2\Phi_0}}{Z^{(D)}_{\text{nz}}} \int_{-\infty}^{\infty} d\xi  \int \left(\prod_{l=1}^\infty \prod_{m=-l}^{m=l} \sqrt{\apr} dx_{lm}\right) e^{-\frac{1}{4 \pi}    \sum_{l=1}^\infty \sum_{m=-l}^{m=l}  l(l+1) e^{\epsilon l(l+1)} x^2_{lm}} \\ &f\left(\sum_{l=1}^\infty    x_{l0} Y_{l0}(0)\right) \ \delta\left(\xi-\sum_{l=1}^\infty    x_{l0} Y_{l0}(0)\right).
\end{align}
The $m\neq0$ terms of the above integral factorises out and cancels the corresponding part in $Z^{(D)}_{\text{nz}}$ and we obtain
\begin{align}
\left\langle f\left(\eta^D(z_0)\right)\right\rangle
&=\frac{e^{-2\Phi_0}}{Z^{(D)}_{\text{nz}(m\neq0)}} \int_{-\infty}^{\infty} d\xi  \int \left(\prod_{l=1}^\infty  \sqrt{\apr} dx_{l0}\right) e^{-\frac{1}{4 \pi}    \sum_{l=1}^\infty   l(l+1) e^{\epsilon l(l+1)} x^2_{l0}} \\ &f\left(\sum_{l=1}^\infty    x_{l0} Y_{l0}(0)\right) \ \delta\left(\xi-\sum_{l=1}^\infty    x_{l0} Y_{l0}(0)\right).
\end{align}
Also, we have $Y_{l0}(0)=\sqrt{\dfrac{2l +1}{4\pi}}=:\omega_l$. So we have
\begin{align}
\left\langle f\left(\eta^D(z_0)\right)\right\rangle
&=\frac{e^{-2\Phi_0}}{Z^{(D)}_{\text{nz}(m\neq0)}} \int_{-\infty}^{\infty} d\xi  \int \left(\prod_{l=1}^\infty  \sqrt{\apr} dx_{l}\right) e^{-\frac{1}{4 \pi}    \sum_{l=1}^\infty   l(l+1) e^{\epsilon l(l+1)} x^2_{l}} \\ &f\left(\sum_{l=1}^\infty  \omega_l  x_{l} \right) \ \delta\left(\xi-\sum_{l=1}^\infty  \omega_l  x_{l} \right),
\end{align}
where we have dropped the subscript $``0"$ from $x_{l0}$ and $Y_{l0}$. We can absorb $\omega_l$ into $x_l$ by a change of integration variable to get
\begin{align}
\left\langle f\left(\eta^D(z_0)\right)\right\rangle
&=\frac{e^{-2\Phi_0}}{Z^{(D)}_{\text{nz}(m\neq0)}} \int_{-\infty}^{\infty} d\xi  \int \left(\prod_{l=1}^\infty  \frac{\sqrt{\apr}}{\omega_l} dx_{l}\right) e^{-\frac{1}{4 \pi}    \sum_{l=1}^\infty   \frac{l(l+1)}{\omega_l^2} e^{\epsilon l(l+1)} x^2_{l}} \\ &f\left(\sum_{l=1}^\infty    x_{l} \right) \ \delta\left(\xi-\sum_{l=1}^\infty    x_{l} \right).
\end{align}

Now we shall pick a mode to integrate over, say $x_1$. Integrating over $x_1$, we obtain
\begin{align}
\left\langle f\left(\eta^D(z_0)\right)\right\rangle
&=\frac{\sqrt{\apr} e^{-2\Phi_0}}{\omega_1 Z^{(D)}_{\text{nz}(m\neq0)}} \int_{-\infty}^{\infty} d\xi \ f\left(\xi\right)  \int \left(\prod_{l=2}^\infty  \frac{\sqrt{\apr}}{\omega_l} dx_{l}\right) \\ & \exp{{-\frac{1}{4 \pi}    \sum_{l=2}^\infty   \frac{l(l+1)}{\omega^2_l} e^{\epsilon l(l+1)} x^2_{l}} -\frac{2}{4 \pi\omega^2_1}        e^{2\epsilon } \left(\xi-\sum_{l=2}^\infty    x_{l}\right)^2} \\
&=\frac{\sqrt{\apr} e^{-2\Phi_0}}{\omega_1 Z^{(D)}_{\text{nz}(m\neq0)}} \int_{-\infty}^{\infty} d\xi \ f\left(\xi\right) e^{-\frac{2}{4 \pi \omega^2_1}        e^{2\epsilon } \xi^2} \int \left(\prod_{l=2}^\infty  \frac{\sqrt{\apr}}{\omega_l} dx_{l}\right) \\ & \exp{-\frac{1}{4 \pi}    \sum_{l=2}^\infty   \frac{l(l+1)}{\omega^2_l} e^{\epsilon l(l+1)} x^2_{l} -\frac{2}{4 \pi\omega^2_1}        e^{2\epsilon } \left(\sum_{l=2}^\infty    x_{l}\right)^2 -\frac{4}{4 \pi\omega^2_1}        e^{2\epsilon } \xi\sum_{l=2}^\infty    x_{l}} \\
&=\frac{\sqrt{\apr} e^{-2\Phi_0}}{\omega_1 Z^{(D)}_{\text{nz}(m\neq0)}} \int_{-\infty}^{\infty} d\xi \ f\left(\xi\right) e^{-\frac{2}{4 \pi\omega^2_1}        e^{2\epsilon } \xi^2} \int \left(\prod_{l=2}^\infty  \frac{\sqrt{\apr}}{\omega_l} dx_{l}\right) e^\Gamma,
\end{align}
where 
\begin{align}
    \Gamma &:= -\frac{1}{4 \pi}    \sum_{l=2}^\infty   \frac{l(l+1)}{\omega^2_l} e^{\epsilon l(l+1)} x^2_{l} -\frac{2}{4 \pi\omega^2_1}        e^{2\epsilon } \left(\sum_{l=2}^\infty    x_{l}\right)^2 -\frac{4}{4 \pi\omega^2_1}        e^{2\epsilon } \xi\sum_{l=2}^\infty    x_{l} \\ &=-\frac{1}{4 \pi}    \sum_{n=2}^\infty \sum_{m=2}^\infty   \frac{n(n+1)}{\omega^2_n} e^{\epsilon n(n+1)} x_{n} \delta_{nm} x_{m} -\frac{2}{4 \pi\omega^2_1}        e^{2\epsilon } \sum_{n=2}^\infty \sum_{m=2}^\infty     x_{n} x_{m}-\frac{4}{4 \pi\omega^2_1}        e^{2\epsilon } \xi\sum_{l=2}^\infty    x_{l}\\
    &=- \sum_{n=2}^\infty \sum_{m=2}^\infty \frac{1}{2 \pi\omega^2_1} e^{2\epsilon }  \left(\frac{\omega^2_1}{\omega^2_n}      \frac{n(n+1)}{2} e^{\epsilon n(n+1)-2\epsilon }\delta_{nm}  +    1      \right) x_{n} x_{m} -\frac{4}{4 \pi\omega^2_1}        e^{2\epsilon } \xi\sum_{l=2}^\infty  x_{l}\\
    &=-   M_{nm} \tilde{x}_{n} \tilde{x}_{m} -\frac{1}{\omega_1}\sqrt{\frac{2}{\pi}}        e^{\epsilon } \xi\sum_{l=2}^\infty  \tilde{x}_{l},
\end{align}
and
\begin{align}
   M_{nm}&:= \left(\tilde{\lambda}_n \delta_{nm}  + 1 \right)\\
   \tilde{\lambda}_n &:=\frac{\omega^2_1}{\omega^2_n}\frac{n(n+1)}{2} e^{\epsilon n(n+1)-2\epsilon } \\
   \tilde{x}_n &:= \frac{1}{\sqrt{2 \pi}\omega_1} e^{\epsilon } x_n
\end{align}

So we get
\begin{align}
\left\langle f\left(\eta^D(z_0)\right)\right\rangle
&=\left(\prod_{l=2}^\infty \frac{\omega_1}{\omega_l} \right) \frac{\sqrt{\apr} e^{-2\Phi_0}}{\omega_1 Z^{(D)}_{\text{nz}(m\neq0)}} \int_{-\infty}^{\infty} d\xi \ f\left(\xi\right) e^{-\frac{1}{2 \pi \omega^2_1}        e^{2\epsilon } \xi^2} \int \left(\prod_{l=2}^\infty  \sqrt{2\pi\apr} e^{-\epsilon }d\tilde{x}_{l}\right) e^\Gamma,
\end{align}

Now let us diagonalise the real symmetric matrix M by an orthogonal transformation $\tilde{x}_n=\Lambda_{nm} y_m$ to get a diagonal matrix $D=\Lambda^T M \Lambda$
\begin{align}
\left\langle f\left(\eta^D(z_0)\right)\right\rangle
&=\left(\prod_{l=2}^\infty \frac{\omega_1}{\omega_l} \right) \frac{\sqrt{\apr} e^{-2\Phi_0}}{\omega_1 Z^{(D)}_{\text{nz}(m\neq0)}} \int_{-\infty}^{\infty} d\xi \ f\left(\xi\right) e^{-\frac{1}{2 \pi \omega^2_1}        e^{2\epsilon } \xi^2}   \frac{1}{\text{Det}(\Lambda)}
\int \left(\prod_{n=2}^\infty  \sqrt{2\pi\apr} e^{-\epsilon }dy_{n}\right) \\ &\exp{- \sum_{n=2}^\infty  D_{n} y^2_n -\xi\sum_{n=2}^\infty k_n y_n},
\end{align}
where $k_n:=\frac{1}{\omega_1} \sum_{m=2}^\infty  \sqrt{\frac{2}{\pi}}        e^{\epsilon } \Lambda_{mn}$ and a factor of $\text{Det}(\Lambda)$ comes from the Jacobian. Now one can complete the squares and shift the mean to get (one can check that all $D_n>0$)
\begin{align}
\left\langle f\left(\eta^D(z_0)\right)\right\rangle
&=\left(\prod_{l=2}^\infty \frac{\omega_1}{\omega_l} \right)\frac{\sqrt{\apr} e^{-2\Phi_0}}{\omega_1 Z^{(D)}_{\text{nz}(m\neq0)}} \int_{-\infty}^{\infty} d\xi \ f\left(\xi\right) e^{-\frac{1}{2 \pi \omega^2_1}        e^{2\epsilon } \xi^2}   \frac{1}{\text{Det}(\Lambda)} \exp{ \sum_{n=2}^\infty \frac{k^2_n}{4D_{n}}\xi^2}
 \\ & \prod_{n=2}^\infty  \int   \sqrt{2\pi\apr} e^{-\epsilon }dy_{n} \exp{-   D_{n} \left(y_n +\xi \frac{k_n}{2D_{n}}\right)^2}\\
 &=\left(\prod_{l=2}^\infty \frac{\omega_1}{\omega_l} \right) \frac{\sqrt{\apr} e^{-2\Phi_0}}{\omega_1 Z^{(D)}_{\text{nz}(m\neq0)}} \frac{1}{\text{Det}(\Lambda)} \frac{1}{\sqrt{\text{Det}(M)}} \left(\prod_{n=2}^\infty     \sqrt{2\apr} \pi e^{-\epsilon }\right) \label{annoyingterms}
 \\ &     \int_{-\infty}^{\infty} d\xi \ f\left(\xi\right) e^{-\frac{1}{2 \pi \omega^2_1}        e^{2\epsilon } \xi^2}    \exp{ \sum_{n=2}^\infty \frac{k^2_n}{4D_{n}}\xi^2}.
\end{align}
All of the annoying factors in \eqref{annoyingterms} gets absorbed into $Z^{(D)}_{\text{nz}(m\neq0)}$ and we finaly get
\begin{align}
\left\langle f\left(\eta^D(z_0)\right)\right\rangle
 &=  \frac{ \int_{-\infty}^{\infty} d\xi \ f\left(\xi\right) e^{-\xi^2/\Omega}}{\int_{-\infty}^{\infty} d\xi \  e^{-\xi^2/\Omega}},\label{coincidentcorrelationformula}
\end{align}
where 
\begin{align}
\Omega^{-1} &:= \frac{e^{2\epsilon}}{2 \pi \omega^2_1}         \left(1 -  \mathbb{M}\right)\,,\\
\mathbb{M}&:=\omega^2_1 \frac{\pi}{2} e^{-2\epsilon} \sum_{n=2}^\infty \frac{k^2_n}{D_{n}} \,.
\end{align}

Now let us calculate $\mathbb{M}$.
\begin{align}
    \mathbb{M}&=\omega^2_1  \frac{\pi}{2} e^{-2\epsilon} \sum_{n=2}^\infty \frac{k^2_n}{D_{n}} = \frac{\pi}{2} e^{-2\epsilon} \sum_{n=2}^\infty \sum_{m=2}^\infty \sum_{\tilde{m}=2}^\infty  \sqrt{\frac{2}{\pi}}        e^{\epsilon } \Lambda_{mn} \frac{1}{D_{n}}  \sqrt{\frac{2}{\pi}}        e^{\epsilon } \Lambda_{\tilde{m}n}\\
    &=\sum_{n=2}^\infty \sum_{m=2}^\infty    [\Lambda D^{-1} \Lambda^T]_{nm} =\sum_{n=2}^\infty \sum_{m=2}^\infty    [M^{-1} ]_{nm},
\end{align}
which follows from $\Lambda$ being an orthogonal matrix. So $\mathbb{M}$ is the sum of all the entries of the inverse of the matrix $M$. One can show that
\begin{align}
    \mathbb{M}= \frac{1}{1+\frac{1}{\sum_{n=2}^\infty \tilde{\lambda}_n^{-1}}}\,\,.
\end{align}

One can also show that 
\begin{align}\label{eq:Omega}
   \Omega=\left( \frac{e^{2\epsilon}}{2 \pi \omega^2_1}  \left(1-\mathbb{M}\right)\right)^{-1} &= \frac{2 \pi \omega^2_1}{e^{2\epsilon}} \left(1+\sum_{n=2}^\infty \tilde{\lambda}_n^{-1}\right)
   = \frac{3}{2}  e^{-2\epsilon}+\sum_{n=2}^\infty \frac{2n+1}{n(n+1)} e^{-\epsilon n(n+1) }\\
   &=\sum_{n=1}^\infty \frac{2n+1}{n(n+1)} e^{-\epsilon n(n+1) } \,.
\end{align}

By Euler–Maclaurin summation formula, $\Omega$ can be written as
\begin{align}\label{eq:Omega_ln_eps}
\Omega = \log \epsilon^{-1} + \dots,
\end{align}
where $\dots$ represent polynomial divergences in $\epsilon$ and also the finite terms.

If we specialize to $f\left(\eta^D(z_0)\right)=\abs{\eta^D(z_0)}^N$, then
the general formula in equation \eqref{coincidentcorrelationformula} reduces to
\begin{align}\label{eq:eta_nth_moment}
\left\langle \abs{\eta^D(z_0)}^N \right\rangle
 &=  \frac{ \int_{-\infty}^{\infty} d\xi \ \abs{\xi}^N e^{-\xi^2/\Omega}}{\int_{-\infty}^{\infty} d\xi \  e^{-\xi^2/\Omega}}= \frac{ \int_{0}^{\infty} d\xi \ \xi^N e^{-\xi^2/\Omega}}{\int_{0}^{\infty} d\xi \  e^{-\xi^2/\Omega}}\\
  &=  \frac{1}{\sqrt{\pi}}  \Gamma\left(\frac{1 + N}{2}\right) \Omega^{N/2}.
\end{align}
We thus get
\begin{align}\label{eq:n-point-corr}
 \left\langle \abs{\eta^D(z_0)}^N \right\rangle
 &=   \frac{1}{\sqrt{\pi}}  \Gamma\left(\frac{1 + N}{2}\right) \left(\sum_{n=1}^\infty \frac{2n+1}{n(n+1)} e^{-\epsilon n(n+1) }\right)^{N/2}\, .
\end{align}

Let us do a consistency check of \eqref{eq:n-point-corr} for $N=2$, in which case, we get
\begin{align}
 \left\langle \eta^D(z_0)\eta^D(z_0) \right\rangle
 &=   \sum_{n=1}^\infty \frac{2n+1}{2n(n+1)} e^{-\epsilon n(n+1) } = \frac{1}{2}\log\epsilon^{-1} + \dots\, ,
\end{align}
which agrees with \cite{Andreev:1990iv}.

\if
The $N=2$ case is a standard result, so we shall compare our result with it. We know that \cite{Andreev:1990iv} 
\begin{align*}\label{eq:Omega_logeps}
\left\langle\eta^\mu(z) \eta^\nu\left(z\right)\right\rangle&=2 \pi  \delta^{\mu \nu}  \left(\frac{1}{4\pi}\sum_{n=1}^\infty \frac{2n+1}{n(n+1)} e^{-\epsilon n(n+1) }\right) \\
&=2 \pi \alpha^{\prime} \delta^{\mu \nu}  \left(-\frac{1}{2\pi} \apr\log \eps  + O(\eps^2) + \text{finite term}\right) \,.
\end{align*}
Thus, we find from \eqref{eq:Omega} that $\Omega = 2 \log \eps^{-1}$. \textcolor{red}{Actually the correct equation is $\Omega =  \log \eps^{-1}$. This mistake arose from the inconsistency acrooss Tseytlins paper.} \RK{Nope!. Whole point of calculating this explicitly is to not rely on the standard know result but to only use it as a benchmark. 
We calculate Omega just from A.53 by Euler maclaurin summation, not by comparing with known result. And then show that it is consistent with the known result. Let’s change this for the next version on ArXiv. I noticed this but missed it later. If we would have taken this shortcut of comparing with known result to find what Omega is, then tedious calculations from A.47 to A.53 would have not been necessary. Calculationally we did a much better job but wrote something that would mislead people in thinking we didn’t do this calculation but just comapred and got Omega instead.}
\fi

Now we focus on the original term we set out to compute: $\left\langle\abs{Y^D+\sqrt{\apr}\eta^D(z_0)}\right\rangle$. For this, the general formula in equation \eqref{coincidentcorrelationformula} reduces to 
\begin{align}
\left\langle \abs{Y^D+\sqrt{\apr}\eta^D(z_0)}\right\rangle
 &=  \frac{ \int_{-\infty}^{\infty} d\xi \ \abs{Y^D+\sqrt{\apr}\xi} e^{-\xi^2/\Omega}}{\int_{-\infty}^{\infty} d\xi \  e^{-\xi^2/\Omega}}=  \frac{1}{\sqrt{\pi \Omega}}  \int_{-\infty}^{\infty} d\xi \ \abs{Y^D+\sqrt{\apr}\xi} e^{-\xi^2/\Omega}\\
 &=\frac{1}{\sqrt{\pi \Omega}}  \int_{-\frac{Y^D}{\sqrt{\apr}}}^{\infty} d\xi \ \left(Y^D+\sqrt{\apr}\xi\right) e^{-\xi^2/\Omega}-\frac{1}{\sqrt{\pi \Omega}}  \int_{-\infty}^{-\frac{Y^D}{\sqrt{\apr}}} d\xi \ \left(Y^D+\sqrt{\apr}\xi\right) e^{-\xi^2/\Omega}\\
 &=\frac{Y^D}{\sqrt{\pi \Omega}}  \int_{-\frac{Y^D}{\sqrt{\apr}}}^{\infty} d\xi \  e^{-\xi^2/\Omega}-\frac{Y^D}{\sqrt{\pi \Omega}}  \int_{-\infty}^{-\frac{Y^D}{\sqrt{\apr}}} d\xi \  e^{-\xi^2/\Omega}\\
 &+\frac{\sqrt{\apr}}{\sqrt{\pi \Omega}}  \int_{-\frac{Y^D}{\sqrt{\apr}}}^{\infty} d\xi \ \xi e^{-\xi^2/\Omega}-\frac{\sqrt{\apr}}{\sqrt{\pi \Omega}}  \int_{-\infty}^{-\frac{Y^D}{\sqrt{\apr}}} d\xi \ \xi e^{-\xi^2/\Omega}\\
 &=\frac{Y^D}{\sqrt{\pi \Omega}}  \frac{\sqrt{\pi \Omega}}{2}\left(1+\text{erf}\left(\frac{Y^D}{\sqrt{\apr \Omega}}\right)\right)-\frac{Y^D}{\sqrt{\pi \Omega}}  \frac{\sqrt{\pi \Omega}}{2}\left(\text{erfc}\left(\frac{Y^D}{\sqrt{\apr \Omega}}\right)\right)\\
 &+\frac{\sqrt{\apr}}{\sqrt{\pi \Omega}}  \frac{\Omega}{2} e^{-\frac{(Y^D)^2}{\apr \Omega}}+\frac{\sqrt{\apr}}{\sqrt{\pi \Omega}}  \frac{\Omega}{2} e^{-\frac{(Y^D)^2}{\apr \Omega}}\\
 &=\sqrt{\Omega}\frac{\sqrt{\apr}}{\sqrt{\pi}}   e^{-\frac{(Y^D)^2}{\apr \Omega}} + \frac{Y^D}{2}  \left(1+\text{erf}\left(\frac{Y^D}{\sqrt{\apr \Omega}}\right)-\text{erfc}\left(\frac{Y^D}{\sqrt{\apr \Omega}}\right)\right)\\
 &=\sqrt{\Omega}\frac{\sqrt{\apr}}{\sqrt{\pi}}   e^{-\frac{(Y^D)^2}{\apr \Omega}} + Y^D  \text{erf}\left(\frac{Y^D}{\sqrt{\apr \Omega}}\right).
\end{align}

So finally, we obtain our desired expression
\begin{align}\label{modoneptexpvalue}
\left\langle \abs{Y^D+\sqrt{\apr}\eta^D(z_0)}\right\rangle
 &=\sqrt{\apr}\left(\frac{\sqrt{\Omega}}{\sqrt{\pi}}   e^{-\frac{(Y^D)^2}{\apr \Omega}} + \frac{Y^D}{\sqrt{\apr}}  \text{erf}\left(\frac{Y^D}{\sqrt{\apr \Omega}}\right)\right) \\
 &= \sqrt{\apr}\left(\left \langle\abs{\eta^D(z_0)}\right\rangle e^{-\frac{(Y^D)^2}{\apr \Omega}} + \frac{Y^D}{\sqrt{\apr}}  \text{erf}\left(\frac{Y^D}{\sqrt{\apr \Omega}}\right)\right) .
\end{align}
where we used \eqref{eq:eta_nth_moment} in the last line that $\left \langle\abs{\eta^D(z_0)}\right\rangle = \frac{\sqrt{\Omega}}{\sqrt{\pi}}$. 

\end{appendices}

\bibliographystyle{JHEP}
\bibliography{main.bib}

@article{Erler:2022agw,
    author = "Erler, Theodore",
    title = "{The closed string field theory action vanishes}",
    eprint = "2204.12863",
    archivePrefix = "arXiv",
    primaryClass = "hep-th",
    doi = "10.1007/JHEP10(2022)055",
    journal = "JHEP",
    volume = "10",
    pages = "055",
    year = "2022"
}

@article{FT1,
    author = "Fradkin, E. S. and Tseytlin, Arkady A.",
    title = "{Quantum String Theory Effective Action}",
    doi = "10.1016/0550-3213(85)90559-0",
    journal = "Nucl. Phys. B",
    volume = "261",
    pages = "1--27",
    year = "1985",
    note = "[Erratum: Nucl.Phys.B 269, 745--745 (1986)]"
}

@article{FT2,
    author = "Fradkin, E. S. and Tseytlin, Arkady A.",
    title = "{Effective Field Theory from Quantized Strings}",
    reportNumber = "LEBEDEV-84-261",
    doi = "10.1016/0370-2693(85)91190-6",
    journal = "Phys. Lett. B",
    volume = "158",
    pages = "316--322",
    year = "1985"
}

@article{FT3,
    author = "Fradkin, E. S. and Tseytlin, Arkady A.",
    title = "{Effective Action Approach to Superstring Theory}",
    reportNumber = "LEBEDEV-85-150",
    doi = "10.1016/0370-2693(85)91468-6",
    journal = "Phys. Lett. B",
    volume = "160",
    pages = "69--76",
    year = "1985"
}

@article{TSEYTLINMobiusInfinitySubtraction1988,
author = "Tseytlin, Arkady A.",
title = "{Mobius Infinity Subtraction and Effective Action in $\sigma$ Model Approach to Closed String Theory}",
reportNumber = "Print-88-0440 (LEBEDEV INST)",
doi = "10.1016/0370-2693(88)90421-2",
journal = "Phys. Lett. B",
volume = "208",
pages = "221--227",
year = "1988"
}

@article{TseytlinZeroMode1989,
    author = "Tseytlin, Arkady A.",
    title = "{Partition Function of String $\sigma$ Model on a Compact Two Space}",
    reportNumber = "Print-89-0041 (CERN)",
    doi = "10.1016/0370-2693(89)90234-7",
    journal = "Phys. Lett. B",
    volume = "223",
    pages = "165--174",
    year = "1989"
}

@article{TseytlinSigmaModelEATachyons2001,
    author = "Tseytlin, Arkady A.",
    title = "{Sigma model approach to string theory effective actions with tachyons}",
    eprint = "hep-th/0011033",
    archivePrefix = "arXiv",
    reportNumber = "OHSTPY-HEP-T-00-025",
    doi = "10.1063/1.1376129",
    journal = "J. Math. Phys.",
    volume = "42",
    pages = "2854--2871",
    year = "2001"
}

@article{TseytlinTachyonEA2001,
    author = "Tseytlin, Arkady A.",
    title = "{Tachyon effective actions in string theory}",
    doi = "10.1023/A:1012372105682",
    journal = "Theor. Math. Phys.",
    volume = "128",
    pages = "1293--1310",
    year = "2001"
}

@article{TseytlinPerelmanEntropy2007,
    author = "Tseytlin, Arkady A.",
    title = "{On sigma model RG flow, 'central charge' action and Perelman's entropy}",
    eprint = "hep-th/0612296",
    archivePrefix = "arXiv",
    reportNumber = "IMPERIAL-TP-AT-6-7",
    doi = "10.1103/PhysRevD.75.064024",
    journal = "Phys. Rev. D",
    volume = "75",
    pages = "064024",
    year = "2007"
}

@article{Erler:2022,
    author = "Erler, Theodore",
    title = "{The closed string field theory action vanishes}",
    eprint = "2204.12863",
    archivePrefix = "arXiv",
    primaryClass = "hep-th",
    month = "4",
    year = "2022"
}

@article{LiuPolchinski1988,
    author = "Liu, Jun and Polchinski, Joseph",
    title = "{Renormalization of the Mobius Volume}",
    reportNumber = "UTTG-26-87",
    doi = "10.1016/0370-2693(88)91566-3",
    journal = "Phys. Lett. B",
    volume = "203",
    pages = "39--43",
    year = "1988"
}

@book{Erbin:2021,
    author = "Erbin, Harold",
    title = "{String Field Theory: A Modern Introduction}",
    doi = "10.1007/978-3-030-65321-7",
    isbn = "978-3-030-65320-0, 978-3-030-65321-7",
    series = "Lecture Notes in Physics",
    volume = "980",
    month = "3",
    year = "2021"
}

@article{RTPhysRevLett.96.181602,
  title = {Holographic Derivation of Entanglement Entropy from the anti--de Sitter Space/Conformal Field Theory Correspondence},
  author = {Ryu, Shinsei and Takayanagi, Tadashi},
  journal = {Phys. Rev. Lett.},
  volume = {96},
  issue = {18},
  pages = {181602},
  numpages = {4},
  year = {2006},
  month = {May},
  publisher = {American Physical Society},
  doi = {10.1103/PhysRevLett.96.181602},
  url = {https://link.aps.org/doi/10.1103/PhysRevLett.96.181602}
}

@article{Kraus:noncompactCFT:2002,
    author = "Kraus, Per and Ryzhov, Anton and Shigemori, Masaki",
    title = "{Strings in noncompact space-times: Boundary terms and conserved charges}",
    eprint = "hep-th/0206080",
    archivePrefix = "arXiv",
    reportNumber = "UCLA-02-TEP-13",
    doi = "10.1103/PhysRevD.66.106001",
    journal = "Phys. Rev. D",
    volume = "66",
    pages = "106001",
    year = "2002"
}

@article{LM2013,
    author = "Lewkowycz, Aitor and Maldacena, Juan",
    title = "{Generalized gravitational entropy}",
    eprint = "1304.4926",
    archivePrefix = "arXiv",
    primaryClass = "hep-th",
    doi = "10.1007/JHEP08(2013)090",
    journal = "JHEP",
    volume = "08",
    pages = "090",
    year = "2013"
}

@article{APS-ClosedStringTachyons-2001,
    author = "Adams, A. and Polchinski, J. and Silverstein, Eva",
    title = "{Don't panic! Closed string tachyons in ALE space-times}",
    eprint = "hep-th/0108075",
    archivePrefix = "arXiv",
    reportNumber = "SLAC-PUB-8955, NSF-ITP-01-75",
    doi = "10.1088/1126-6708/2001/10/029",
    journal = "JHEP",
    volume = "10",
    pages = "029",
    year = "2001"
}

@article{KT:2001,
    author = "Kazakov, V. A. and Tseytlin, Arkady A.",
    title = "{On free energy of 2-D black hole in bosonic string theory}",
    eprint = "hep-th/0104138",
    archivePrefix = "arXiv",
    reportNumber = "LPTENS-01-22, OHSTPY-HEP-T-01-008",
    doi = "10.1088/1126-6708/2001/06/021",
    journal = "JHEP",
    volume = "06",
    pages = "021",
    year = "2001"
}

@phdthesis{Mertens:Thesis:2015,
    author = "Mertens, Thomas G.",
    title = "{Hagedorn String Thermodynamics in Curved Spacetimes and near Black Hole Horizons}",
    eprint = "1506.07798",
    archivePrefix = "arXiv",
    primaryClass = "hep-th",
    school = "Gent U.",
    year = "2015"
}

@article{Eberhardt:diskPF:2021,
    author = "Eberhardt, Lorenz and Pal, Sridip",
    title = "{The disk partition function in string theory}",
    eprint = "2105.08726",
    archivePrefix = "arXiv",
    primaryClass = "hep-th",
    doi = "10.1007/JHEP08(2021)026",
    journal = "JHEP",
    volume = "08",
    pages = "026",
    year = "2021"
}

@article{Jafferis-ER=EPR:2021,
    author = "Jafferis, Daniel L. and Schneider, Elliot",
    title = "{Stringy ER=EPR}",
    eprint = "2104.07233",
    archivePrefix = "arXiv",
    primaryClass = "hep-th",
    month = "4",
    year = "2021"
}

@article{silverstein2023black,
  title="Black hole to cosmic horizon microstates in string/M theory: timelike boundaries and internal averaging",
  author={Silverstein, Eva},
  journal={Journal of High Energy Physics},
  volume={2023},
  number={5},
  pages={1--17},
  year={2023},
  publisher={Springer},
  eprint="2212.00588",
  archivePrefix = "arXiv",
}

@article{troost2011ads3,
  title={The AdS3 central charge in string theory},
  author={Troost, Jan},
  journal={Physics Letters B},
  volume={705},
  number={3},
  pages={260--263},
  year={2011},
  publisher={Elsevier}
}

@article{eberhardt2023holographic,
  title={Holographic Weyl anomaly in string theory},
  author={Eberhardt, Lorenz and Pal, Sridip},
  journal={arXiv preprint arXiv:2307.03000},
  year={2023}
}

@article{grebenkov2019probability,
  title={Probability distribution of the boundary local time of reflected Brownian motion in Euclidean domains},
  author={Grebenkov, Denis S},
  journal={Physical Review E},
  volume={100},
  number={6},
  pages={062110},
  year={2019},
  publisher={APS}
}

@article{björk2015pedestrians,
      title={The Pedestrian's Guide to Local Time}, 
      author={Tomas Björk},
      year={2015},
      eprint={1512.08912},
      archivePrefix={arXiv},
      primaryClass={math.PR}
}

@misc{random,
  author = {Kyle Siegrist},
  title = {Probability, Mathematical Statistics, Stochastic Processes},
  howpublished = "\url{https://www.randomservices.org/random/special/FoldedNormal.html/}",
}

@article{wallner2020half,
  title={A half-normal distribution scheme for generating functions},
  author={Wallner, Michael},
  journal={European Journal of Combinatorics},
  volume={87},
  pages={103138},
  year={2020},
  publisher={Elsevier}
}

@article{banderier2018local,
  title={Local time for lattice paths and the associated limit laws},
  author={Banderier, Cyril and Wallner, Michael},
  journal={arXiv preprint arXiv:1805.09065},
  year={2018}
}

@misc{HNDMathWorks,
  author = {Weisstein, Eric W.},
  title = {Half-Normal Distribution." From MathWorld--A Wolfram Web Resource},
  howpublished = "\url{  https://mathworld.wolfram.com/Half-NormalDistribution.html}",
}

@book{morters2010brownian,
  title={Brownian motion},
  author={M{\"o}rters, Peter and Peres, Yuval},
  volume={30},
  year={2010},
  publisher={Cambridge University Press}
}

@book{karatzas2014brownian,
  title={Brownian motion and stochastic calculus},
  author={Karatzas, Ioannis and Shreve, Steven},
  volume={113},
  year={2014},
  publisher={springer}
}

@book{borodin2015handbook,
  title={Handbook of Brownian motion-facts and formulae},
  author={Borodin, Andrei N and Salminen, Paavo},
  year={2015},
  publisher={Springer Science \& Business Media}
}

@article{skorokhod1961stochastic,
  title={Stochastic equations for diffusion processes in a bounded region},
  author={Skorokhod, Anatoliy V},
  journal={Theory of Probability \& Its Applications},
  volume={6},
  number={3},
  pages={264--274},
  year={1961},
  publisher={SIAM}
}

@article{skorokhod1962stochastic,
  title={Stochastic equations for diffusion processes in a bounded region. II},
  author={Skorokhod, Anatoliy V},
  journal={Theory of Probability \& Its Applications},
  volume={7},
  number={1},
  pages={3--23},
  year={1962},
  publisher={SIAM}
}

@article{tanaka1979stochastic,
  title={Stochastic differential equations with reflecting},
  author={Tanaka, Hiroshi},
  journal={Stochastic Processes: Selected Papers of Hiroshi Tanaka},
  volume={9},
  pages={157},
  year={1979}
}

@article{it1965diffusion,
  title={Diffusion processes and their sample paths},
  author={Ito, Kiyosi and McKean, HP},
  journal={Die Grundlehren der Mathematischen Wissenschaften in Einzeldarstellungen},
  volume={125},
  year={1965}
}

@inproceedings{levy1965processus,
  title={Processus Stochastiques et Mouvement Brownien (Paris: Gauthier-Villars) Google Scholar L{\'e}vy P 1951},
  author={L{\'e}vy, P},
  booktitle={Proc. 2nd Berkeley Symp. on Mathematical Statistics and Probability},
  year={1965}
}

@article{takacs1995local,
  title={On the local time of the Brownian motion},
  author={Tak{\'a}cs, Lajos},
  journal={The Annals of Applied Probability},
  pages={741--756},
  year={1995},
  publisher={JSTOR}
}

@article{Clark1980half,
  title = {Quantum mechanics on the half-line using path integrals},
  author = {Clark, T. E. and Menikoff, R. and Sharp, D. H.},
  journal = {Phys. Rev. D},
  volume = {22},
  issue = {12},
  pages = {3012--3016},
  numpages = {0},
  year = {1980},
  month = {Dec},
  publisher = {American Physical Society},
  doi = {10.1103/PhysRevD.22.3012},
  url = {https://link.aps.org/doi/10.1103/PhysRevD.22.3012}
}

@book{mckean1969stochastic,
  title={Stochastic integrals},
  author={McKean, Henry P},
  volume={353},
  year={1969},
  publisher={American Mathematical Soc.}
}

@article{Halder:2023adw,
    author = "Halder, Indranil and Jafferis, Daniel L.",
    title = "{Thermal Bekenstein-Hawking entropy from the worldsheet}",
    eprint = "2310.02313",
    archivePrefix = "arXiv",
    primaryClass = "hep-th",
    month = "10",
    year = "2023"
}

@article{Mahajan:2021nsd,
    author = "Mahajan, Raghu and Stanford, Douglas and Yan, Cynthia",
    title = "{Sphere and disk partition functions in Liouville and in matrix integrals}",
    eprint = "2107.01172",
    archivePrefix = "arXiv",
    primaryClass = "hep-th",
    doi = "10.1007/JHEP07(2022)132",
    journal = "JHEP",
    volume = "07",
    pages = "132",
    year = "2022"
}

@article{Eberhardt:2023lwd,
    author = "Eberhardt, Lorenz and Pal, Sridip",
    title = "{Holographic Weyl anomaly in string theory}",
    eprint = "2307.03000",
    archivePrefix = "arXiv",
    primaryClass = "hep-th",
    doi = "10.21468/SciPostPhys.16.1.027",
    journal = "SciPost Phys.",
    volume = "16",
    number = "1",
    pages = "027",
    year = "2024"
}

@article{Bastianelli:2006hq,
    author = "Bastianelli, Fiorenzo and Corradini, Olindo and Pisani, Pablo A. G.",
    title = "{Worldline approach to quantum field theories on flat manifolds with boundaries}",
    eprint = "hep-th/0612236",
    archivePrefix = "arXiv",
    doi = "10.1088/1126-6708/2007/02/059",
    journal = "JHEP",
    volume = "02",
    pages = "059",
    year = "2007"
}

@article{Bastianelli:2008vh,
    author = "Bastianelli, Fiorenzo and Corradini, Olindo and Pisani, Pablo A. G. and Schubert, Christian",
    title = "{Scalar heat kernel with boundary in the worldline formalism}",
    eprint = "0809.0652",
    archivePrefix = "arXiv",
    primaryClass = "hep-th",
    reportNumber = "AEI-2008-054, UMSNH-IFM-F-2008-25",
    doi = "10.1088/1126-6708/2008/10/095",
    journal = "JHEP",
    volume = "10",
    pages = "095",
    year = "2008"
}

@inproceedings{Bastianelli:2009mw,
    author = "Bastianelli, F. and Corradini, O. and Pisani, P. A. G. and Schubert, C.",
    title = "{Worldline Approach to QFT on Manifolds with Boundary}",
    booktitle = "{9th Conference on Quantum Field Theory under the Influence of External Conditions (QFEXT 09): Devoted to the Centenary of H. B. G. Casimir}",
    eprint = "0912.4120",
    archivePrefix = "arXiv",
    primaryClass = "hep-th",
    doi = "10.1142/9789814289931_0051",
    pages = "415--420",
    year = "2010"
}

@article{Andreev:1990iv,
    author = "Andreev, O. D. and Metsaev, R. R. and Tseytlin, Arkady A.",
    title = "{Covariant calculation of the statistical sum of the two-dimensional sigma model on compact two surfaces.}",
    eprint = "2301.02867",
    archivePrefix = "arXiv",
    primaryClass = "hep-th",
    journal = "Sov. J. Nucl. Phys.",
    volume = "51",
    pages = "359--366",
    year = "1990"
}

@article{Muhlmann:2021clm,
    author = {M\"uhlmann, Beatrix},
    title = "{The two-sphere partition function in two-dimensional quantum gravity at fixed area}",
    eprint = "2106.04532",
    archivePrefix = "arXiv",
    primaryClass = "hep-th",
    doi = "10.1007/JHEP09(2021)189",
    journal = "JHEP",
    volume = "09",
    number = "189",
    pages = "189",
    year = "2021"
}

@article{Araujo-Regado:2022gvw,
    author = "Araujo-Regado, Goncalo and Khan, Rifath and Wall, Aron C.",
    title = "{Cauchy slice holography: a new AdS/CFT dictionary}",
    eprint = "2204.00591",
    archivePrefix = "arXiv",
    primaryClass = "hep-th",
    doi = "10.1007/JHEP03(2023)026",
    journal = "JHEP",
    volume = "03",
    pages = "026",
    year = "2023"
}

@article{Batra:2024kjl,
    author = "Batra, Gauri and De Luca, G. Bruno and Silverstein, Eva and Torroba, Gonzalo and Yang, Sungyeon",
    title = "{Bulk-local dS$_3$ holography: the Matter with $T\bar T+\Lambda_2$}",
    eprint = "2403.01040",
    archivePrefix = "arXiv",
    primaryClass = "hep-th",
    month = "3",
    year = "2024"
}

@article{GH,
  title = {Action integrals and partition functions in quantum gravity},
  author = {Gibbons, G. W. and Hawking, S. W.},
  journal = {Phys. Rev. D},
  volume = {15},
  issue = {10},
  pages = {2752--2756},
  numpages = {0},
  year = {1977},
  month = {May},
  publisher = {American Physical Society},
  doi = {10.1103/PhysRevD.15.2752},
  url = {https://link.aps.org/doi/10.1103/PhysRevD.15.2752}
}

@article{York:1972sj,
    author = "York, Jr., James W.",
    title = "{Role of conformal three geometry in the dynamics of gravitation}",
    doi = "10.1103/PhysRevLett.28.1082",
    journal = "Phys. Rev. Lett.",
    volume = "28",
    pages = "1082--1085",
    year = "1972"
}

@article{Einstein:1916cd,
    author = "Einstein, Albert",
    title = "{Hamilton's Principle and the General Theory of Relativity}",
    journal = "Sitzungsber. Preuss. Akad. Wiss. Berlin (Math. Phys. )",
    volume = "1916",
    pages = "1111--1116",
    year = "1916"
}

@article{Cavaglia:2016oda,
    author = "Cavagli\`a, Andrea and Negro, Stefano and Sz\'ecs\'enyi, Istv\'an M. and Tateo, Roberto",
    title = "{$T \bar{T}$-deformed 2D Quantum Field Theories}",
    eprint = "1608.05534",
    archivePrefix = "arXiv",
    primaryClass = "hep-th",
    doi = "10.1007/JHEP10(2016)112",
    journal = "JHEP",
    volume = "10",
    pages = "112",
    year = "2016"
}

@article{McGough:2016lol,
    author = "McGough, Lauren and Mezei, M\'ark and Verlinde, Herman",
    title = "{Moving the CFT into the bulk with $ T\overline{T} $}",
    eprint = "1611.03470",
    archivePrefix = "arXiv",
    primaryClass = "hep-th",
    doi = "10.1007/JHEP04(2018)010",
    journal = "JHEP",
    volume = "04",
    pages = "010",
    year = "2018"
}

@article{Smirnov:2016lqw,
    author = "Smirnov, F. A. and Zamolodchikov, A. B.",
    title = "{On space of integrable quantum field theories}",
    eprint = "1608.05499",
    archivePrefix = "arXiv",
    primaryClass = "hep-th",
    doi = "10.1016/j.nuclphysb.2016.12.014",
    journal = "Nucl. Phys. B",
    volume = "915",
    pages = "363--383",
    year = "2017"
}

@article{Zamolodchikov:2004ce,
    author = "Zamolodchikov, Alexander B.",
    title = "{Expectation value of composite field T anti-T in two-dimensional quantum field theory}",
    eprint = "hep-th/0401146",
    archivePrefix = "arXiv",
    reportNumber = "BONN-TH-2004-02",
    month = "1",
    year = "2004"
}

@article{Sen:2024nfd,
    author = "Sen, Ashoke and Zwiebach, Barton",
    title = "{String Field Theory: A Review}",
    eprint = "2405.19421",
    archivePrefix = "arXiv",
    primaryClass = "hep-th",
    reportNumber = "MIT-CTP/5653",
    month = "5",
    year = "2024"
}

@article{Gibbons:1976ue,
    author = "Gibbons, G. W. and Hawking, S. W.",
    title = "{Action Integrals and Partition Functions in Quantum Gravity}",
    reportNumber = "PRINT-76-0995 (CAMBRIDGE)",
    doi = "10.1103/PhysRevD.15.2752",
    journal = "Phys. Rev. D",
    volume = "15",
    pages = "2752--2756",
    year = "1977"
}

@article{Ahmadain:2022eso,
    author = "Ahmadain, Amr and Wall, Aron C.",
    title = "{Off-Shell Strings II: Black Hole Entropy}",
    eprint = "2211.16448",
    archivePrefix = "arXiv",
    primaryClass = "hep-th",
    month = "11",
    year = "2022"
}

@article{Ahmadain:2022tew,
    author = "Ahmadain, Amr and Wall, Aron C.",
    title = "{Off-Shell Strings I: S-matrix and Action}",
    eprint = "2211.08607",
    archivePrefix = "arXiv",
    primaryClass = "hep-th",
    month = "11",
    year = "2022"
}

@article{Firat:2024kxq,
    author = "F\i{}rat, Atakan Hilmi and Mamade, Raji Ashenafi",
    title = "{Boundary terms in string field theory}",
    eprint = "2411.16673",
    archivePrefix = "arXiv",
    primaryClass = "hep-th",
    reportNumber = "MIT-CTP-5810",
    month = "11",
    year = "2024"
}

@article{Ahmadain:2024hdp,
    author = "Ahmadain, Amr and Frenkel, Alexander and Wall, Aron C.",
    title = "{A Background-Independent Closed String Action at Tree Level}",
    eprint = "2410.11938",
    archivePrefix = "arXiv",
    primaryClass = "hep-th",
    month = "10",
    year = "2024"
}

@article{Ahmadain:2024uyo,
    author = "Ahmadain, Amr and Akhtar, Shoaib and Khan, Rifath",
    title = "{The GHY boundary term from the string worldsheet to linear order}",
    eprint = "2411.06400",
    archivePrefix = "arXiv",
    primaryClass = "hep-th",
    month = "11",
    year = "2024"
}

@article{Ahmadain:2022gfw,
    author = "Ahmadain, Amr and Frenkel, Alexander and Ray, Krishnendu and Soni, Ronak M.",
    title = "{Boundary description of microstates of the two-dimensional black hole}",
    eprint = "2210.11493",
    archivePrefix = "arXiv",
    primaryClass = "hep-th",
    doi = "10.21468/SciPostPhys.16.1.020",
    journal = "SciPost Phys.",
    volume = "16",
    number = "1",
    pages = "020",
    year = "2024"
}

\end{document}